\documentclass[useAMS,usenatbib]{mn2e}

\include{epsf}
\usepackage{graphicx}
\usepackage{times}

\newcommand{\hi}{H\,{\sc i}}
\newcommand{\hii}{H\,{\sc ii}}
\newcommand{\mgii}{Mg\,{\sc ii}}

\newcommand{\msol}{\mbox{${\rm M}_\odot$}}
\newcommand{\hubble}{\mbox{$\rm km\, s^{-1}\, Mpc^{-1}$}}
\newcommand{\kms}{\mbox{$\rm km\, s^{-1}$}}

\newcommand{\mhi}{\mbox{$M_{\rm HI}$}}

\newcommand{\nhi}{\mbox{$N_{\rm HI}$}}
\newcommand{\fnhi}{\mbox{$f(N_{\rm HI})$}}

\newcommand{\nhis}{\mbox{$N^*_{\rm HI}$}}
\newcommand{\ohi}{\mbox{$\Omega_{\rm HI}$}}
\newcommand{\mhis}{\mbox{$M^*_{\rm HI}$}}

\newcommand{\icmsq}{\mbox{$ \rm cm^{-2}$}}
\newcommand{\redshift}{\mbox{$z$}}

\title[DLAs and the local galaxy population]{Reconciling the local
galaxy population with damped Ly $\alpha$ cross-sections and metal
abundances} 

\author[M.~A. Zwaan et al.]{M.~A. Zwaan,$^1$\thanks{E-mail: mzwaan@eso.org} 
	J.~M. van der Hulst,$^2$ 
	F.~H. Briggs,$^{3,4}$ 
	M.~A.~W. Verheijen$^2$ and \newauthor
	E.~V. Ryan-Weber$^5$ \\
     $^1$ European Southern Observatory, Karl-Schwarzschild-Str. 2,
     85748 Garching b. M{\"u}nchen, Germany.\\
     $^2$ Kapteyn Astronomical Institute, PO box 800, 9700 AV
     Groningen, The Netherlands\\
     $^3$ Research School for Astronomy \& Astrophysics, Mount Stromlo
     Observatory, Cotter Road, Weston, ACT 2611, Australia\\
     $^4$ Australian National Telescope Facility, PO Box 76, Epping,
     NSW 1710, Australia\\
     $^5$ Institute of Astronomy, University of Cambridge, Madingley
     Road, Cambridge CB30HA, UK }

\begin{document}

\date{Accepted ...
      Received ...}

\pagerange{\pageref{firstpage}--\pageref{lastpage}}
\pubyear{0000}

\maketitle

\label{firstpage}

\begin{abstract}
A comprehensive analysis of 355 high-quality WSRT \hi\ 21-cm line maps
of nearby galaxies shows that the properties and incident rate of
Damped Lyman $\alpha$ absorption systems (DLAs) observed in the spectra
of high redshift QSOs are in good agreement with DLAs originating in
gas disks of galaxies like those in the $z\approx 0$ population.
Comparison of low $z$ DLA statistics with the \hi\ incidence rate and
column density distribution $\fnhi$ for the local galaxy sample shows
no evidence for evolution in the integral ``cross section density''
$<n\sigma>=l^{-1}$ ($l$=mean free path between absorbers) below
$z\approx 1.5$, implying that there is no need for a hidden population
of galaxies or \hi\ clouds to contribute significantly to the DLA
cross section.  Compared with $z\approx 4$, our data indicates
evolution of a factor of two in the comoving density along a line of
sight.  We find that $dN/dz(z=0)=0.045 \pm 0.006$.  The idea that the
local galaxy population can explain the DLAs is further strengthened
by comparing the properties of DLAs and DLA galaxies with the
expectations based on our analysis of local galaxies.  The
distribution of luminosities of DLA host galaxies, and of impact
parameters between QSOs and the centres of DLA galaxies, are in good
agreement with what is expected from local galaxies.  Approximately
87 per cent of low $z$ DLA galaxies are expected to be fainter than $L_*$
and 37 per cent have impact parameters less than 1$''$ at $z=0.5$. The
analysis shows that some host galaxies with very low impact parameters
and low luminosities are expected to be missed in optical follow up
surveys.  The well-known metallicity--luminosity relation in galaxies,
in combination with metallicity gradients in galaxy disks, cause the
expected median metallicity of low redshift DLAs to be low ($\sim$1/7
solar), which is also in good agreement with observations of low $z$
DLAs.  We find that \fnhi\ can be fitted satisfactorily with a
gamma distribution, a single power-law is not a good fit at the
highest column densities $\nhi>10^{21}~\icmsq$.  The vast majority
($\approx 81$ per cent) of the \hi\ gas in the local Universe resides in
column densities above the classical DLA limit ($\nhi>2\times
10^{20}\icmsq$), with $\nhi \sim 10^{21}\icmsq$ dominating the cosmic
\hi\ mass density.

\end{abstract}
 

\section{Introduction}
How does the cosmological mass density of neutral hydrogen (\hi) gas
evolve over the history of the Universe and what sort of galaxies are
responsible for this evolution?  Two completely unrelated
observational techniques are used to find answers to these questions
for the present epoch and at earlier cosmic times.  At intermediate
and high redshifts (out to $z\sim6$) deep optical and UV spectra of
background quasars are scrutinised to find high column density \hi\
that causes optical depth $\tau_{\rm LL}\ga 1$ blueward of the Lyman
limit. In detecting these absorbing systems, their distance is not a
limiting factor since the detection depends only on the brightness of
the background source against which the absorber is found. However,
the nature of the absorbing system is difficult to determine because
the absorption spectrum only gives information along a very narrow
sight-line through the system. Despite the large effort that has been
dedicated to identify and characterise the high redshift absorbers through deep
optical imaging, very little about their properties is yet known
\citep[e.g.,][and references therein]{moller2002}.

At $z=0$, 21-cm emission line surveys are used to study the high
column density \hi. Blind surveys with the Arecibo and Parkes radio
telescopes have resulted in accurate measurements of the cosmic \hi\ mass density 
$\ohi(z=0)$
\citep{zwaan1997, rosenberg2002, zwaan2003, zwaan2005}. In contrast to
the high redshift studies, the 21-cm emission line surveys readily
result in measurements of the total \hi\ masses of the detected
objects. Furthermore, after follow-up with an aperture synthesis
instrument, the 3-dimensional \hi\ data cube can be obtained, from
which a detailed velocity field and column density map can be
derived. Unfortunately, the \hi\ 21-cm hyperfine line is very weak,
which limits the studies to the very local Universe ($z<0.2$). The
column densities that routine 21-cm emission line observations can
sense, are the same as those of the highest column density Ly$\alpha$
absorbers known as Damped Ly$\alpha$ systems, or DLAs, with column
densities $\nhi>2\times 10^{20}\, \icmsq$. These are the systems that
contain most of the cosmic \hi\ mass 
\citep[e.g.,][]{wolfe1986, lanzetta1991,storrie-lombardi2000},
although there are indications that systems with slightly lower column
densities may contribute approximately 20 per cent to \ohi\
\citep{peroux2005}.

The purpose of this paper is to test whether the results from the
$z=0$ 21-cm line surveys and the $z>0$ UV and optical surveys can be
reconciled. This question relates directly to the problem of the
origin of DLA systems. Traditionally, DLAs were thought to arise in
large gaseous disks in the process of evolving to present-day spiral
galaxies \citep{wolfe1986, wolfe1995}. This idea was supported by the
fact that the velocity profiles of unsaturated, low-ion metal lines
are consistent with rapidly rotating, large thick disks
\citep{prochaska1997,prochaska1998}. Dissenting views do exist,
perhaps most notably presented by \citet{haehnelt1998}, who argued
that the velocity structure can alternatively be explained by
protogalactic clumps coalescing on dark matter haloes \citep[see also
][]{ledoux1998, khersonsky1996}. In this view, the DLA systems are
aggregates of dense clouds with complex kinematics rather than ordered
rotating gas disks.

Imaging surveys for DLA host galaxies have so far not resulted in a
consistent picture of their properties \citep{lebrun1997,fynbo1999,
kulkarni2000, bouche2001, warren2001, colbert2002, moller2002,
prochaska2002, rao2003, lacy2003, chen2003}.  The few successful
identifications show that a mixed population of galaxies is
responsible for the DLA cross section. Semi-analytical and numerical
models of galaxy formation point to sub-$L_*$ galaxies as the major
contributors to \hi\ cross section above the DLA limit
\citep[e.g.,][]{kauffmann1996, gardner2001, nagamine2004, okoshi2005}.

In this paper we concentrate primarily on cross-section
arguments. \citet{burbidge1977} started the discussion on whether the
disks of normal galaxies can explain the incidence rate of absorbers.
The method consists of multiplying the space density of galaxies
(measured through the optical luminosity function) with the average
area of the \hi\ disk above the DLA column density
limit. \citet{burbidge1977} concluded that the mean free path for
interception with a galactic disk was too large to account for the
number of observed absorbers.  \citet{wolfe1986},
\citet{lanzetta1991}, \citet{fynbo1999}, \citet{schaye2001}, and
\citet{chen2003} use variations of this same simple analytical
approach for estimates at high and low $z$.  Although these approaches
are useful to gain insight in the problem, a much more direct method
is to use real 21-cm column density maps of local galaxies to evaluate
the DLA incidence rate.  This method was first used by \citet{rao1993}
who used Gaussian fits to the radial \hi\ distribution of 27 galaxies
as measured with the Arecibo Telescope. Much higher resolution \hi\
maps were used by \citet{zwaan2002} who used WSRT maps of a
volume-limited sample of 49 galaxies in the Ursa Major cluster
\citep{verheijen2001}. A similar study was conducted by
\citet{ryan-weber2003} based on 35 galaxies selected from HIPASS
\citep{meyer2004}, which were observed with the ATCA. Finally,
\citet{rosenberg2003} used low-resolution ($\sim 45''$) VLA maps of
their sample of 50 \hi-selected galaxies.

With the completion of WHISP (The Westerbork \hi\ survey of Spiral and
Irregular galaxies, \citealt{vanderhulst2001}) we have available a
much larger sample of galaxies, with a high dynamic range in galaxy
properties and observed with a spatial resolution of $\approx
12''\times 12''/\sin\delta$. With this sample, the cross section
analysis can be done more accurately and in more detail.  This paper
is organised as follows.  In the following section we describe the
details of the WHISP sample, in section 3 we summarise surveys for low
redshift DLA host galaxies and compile a sample of these systems that
we use for our comparison with $z=0$.  We calculate the \hi\ column
density distribution function in section 4 and compare the results
with the statistics at higher redshifts in section 5. A comparison of
the redshift number density $dN/dz$ between $z=0$ and higher redshift
is given in section 6. In section 7 we calculate the probability
distribution functions of various parameters of cross-section selected
galaxy samples and show that these agree well with results of low-$z$
DLA host galaxy searches.  In Section 8 we show that the observed
metallicities in DLAs are in good agreement with the expected values
if low-$z$ DLAs arise in the gas disks of galaxies typical of the
$z=0$ population.  Section 9 summarises the results.  We use
$H_0=75~\hubble$ throughout this paper to calculate distance dependent
quantities.


\section{The WHISP sample}\label{whisp.sec}
\begin{figure*}
\centering
\includegraphics[width=18cm,trim=0.5cm 4.6cm 0.2cm 3cm ]{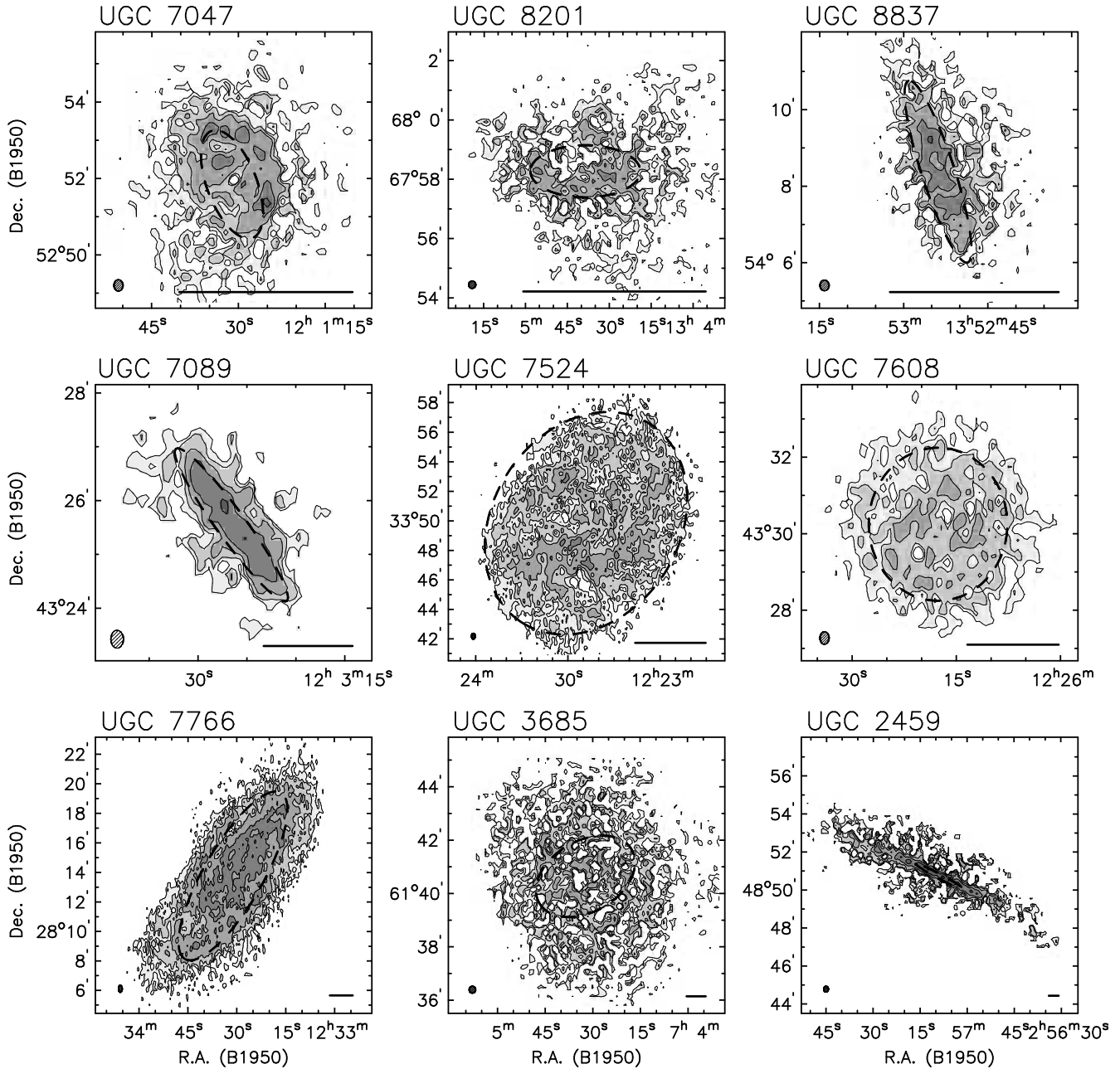}
    \caption{\hi\ column density maps of six of the WHISP
    galaxies. The top three galaxies have total \hi\ masses of
    $10^8\msol$, the middle three have $\mhi\sim 10^{9}\msol$, and the
    bottom three have $\mhi\sim 10^{10}\msol$. Contour levels are at
    column densities of $0.2$, $0.5$, $1$, $2$, $4$, $8$ and $16\times
    10^{21}\, \icmsq$, where the lowest contour corresponds to the DLA
    limit.  The beam sizes are indicated in the lower left corners of
    each panel. The ellipses indicate the orientation parameters of
    the optical images of these galaxies and the extent corresponds to
    the diameter measured at the 25th $B$-magnitude isophote.  The
    scale bar in the lower right corner corresponds to 5 kpc at the
    distance of the galaxies.  }
    \label{examplemaps.fig}
\vspace{-0.25cm}\end{figure*}

The incidence rate of high column density \hi\ in the local Universe
can best be calculated from 21-cm maps of a large sample of galaxies
with uniform selection criteria and high spatial resolution. Such a
sample is available through WHISP (The Westerbork \hi\ survey of
Spiral and Irregular galaxies, \citealt{vanderhulst2001}), an \hi\
survey of some 390 galaxies carried out with the Westerbork Synthesis
Radio Telescope in the periods 1992 to 1998 and 2000 to 2002.  The aim
of this survey is to obtain maps of the distribution and velocity
structure of the \hi\ in a large sample of galaxies, covering all
Hubble types from S0 to Im and a considerable range in
luminosity. This survey increases the number of well-analysed \hi\
observations of galaxies by an order of magnitude, with the principal
goal of investigating the systematics of rotation curves and the
properties of the dark matter haloes they trace.  First results on the
late type galaxies have been published \citep{swaters2002} and a study
of the early type galaxies is on its way
\citep{noordermeer2004,noordermeer2005}.  Examples of \hi\ maps of nine
galaxies used in our analysis are shown in
Fig.~\ref{examplemaps.fig}, where the top three galaxies have \hi\
masses of $\mhi\sim 10^{8}\msol$, the middle three have $\mhi\sim
10^{9}\msol$, and the bottom three have $\mhi\sim 10^{10}\msol$.

The sample from which WHISP candidates have been selected consists of
galaxies in the Uppsala General Catalogue (UGC) of Galaxies
\citep{nilson1973} that fit the following criteria: 1) $B$-band major
axis diameter more extended than $1.2 \arcmin$; 2) declination (B1950)
north of $+20^{\circ}$; and 3) the 21-cm flux density averaged over
the \hi\ profile exceeds $100$ mJy. For observations after the year
2000 the flux density criterion was relaxed to also include galaxies
with lower average flux densities. This enabled observation of a
reasonably large number of galaxies of Hubble type S0 - Sb,
which typically have lower \hi\ fluxes than galaxies of Hubble type
Sbc and later. For the analysis presented in this paper we made use of
the data of 355 galaxies out of the total of 391 in the WHISP sample,
which had been fully reduced at the time this analysis started.  The
distribution over Hubble type and optical luminosity for the 355
galaxies in our sample is given in table~1.

\begin{table*}
\centering
\begin{minipage}{120mm}
\caption{Properties of the WHISP sample}
\label{relcon.tab}
\begin{tabular}{l c c c c c}
\hline\hline
   & $L_B<L_B^*/20$ & $ L_B^*/20< L_B<L_B^*/5 $ & $L_B^*/5< L_B<L_B^* $ & $L_B>L_B^*$ & Total \\
\hline
E/S0 &     2           &     2           &     8           &     9           &    21\\
Sa/Sb &     4           &     7           &    31           &    41           &    83\\
Sbc/Sc &     7           &    14           &    37           &    52           &   110\\
Scd/Sd &    19           &    15           &    17           &     4           &    55\\
Sm/Irr &    43           &    31           &     7           &     5           &    86\\
Total &    75           &    69           &   100           &   111           &   355\\
\hline
\end{tabular}
\end{minipage}
\end{table*}

A total of about 4500 hours of observing time (more than one year of
continuous data taking) was used over a period of 10 years to observe
391 galaxies. Of these, 281 were observed with the old WSRT system
between 1992 and 1998. The remaining 110 galaxies were observed with
the upgraded, more sensitive system between 2000 and 2002.  The main
difference between the two periods is the upgrade of the WSRT with
cooled receivers on all telescopes and a greatly improved bandwidth
and correlator capacity.  In practice this means that the sensitivity
of the observations made in 2000 and thereafter are a factor of $\sim
4$ better than before (e.g.  a typical r.m.s. of 0.8 mJy/beam as
compared to 3 mJy/beam).  The velocity resolution for most galaxies is
5 km s$^{-1}$. The highly inclined galaxies in the sample with
inclinations typically above $ 60 ^{\circ}$ have been observed at a
lower velocity resolution of 20 km s$^{-1}$, while more face-on
galaxies and galaxies with narrow single dish profiles ($\Delta v \le
60\, \kms$) have been observed with a velocity resolution of $2.5
\,\kms$.  

The WSRT observations of the WHISP galaxies have been processed
following a well described standard procedure consisting of a (u,v)
data processing phase (data editing, calibration and Fourier
transformation) using the native WSRT reduction package NewStar and an
image processing phase (continuum subtraction, {\sc clean}ing,
determination of \hi\ distributions and velocity fields using GIPSY.
The resulting data cubes and images have been archived and a summary
is displayed on the web at {\it
http://www.astro.rug.nl/$\sim$whisp}. A precise description of the
reduction procedures can also be found there.

\begin{figure}\centering
\includegraphics[width=\columnwidth,trim=0cm 10.0cm 0cm 0cm]{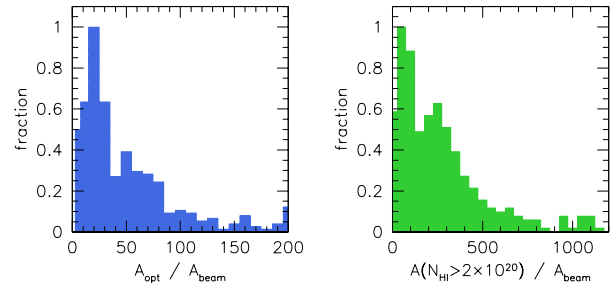}
    \caption{{\em Left\/:} Distribution of number of beams over the
     optical area within $D_{25}$.  {\em Right\/:} Distribution of
     number of beams over the area with \hi\ column density exceeding
     the DLA limit of $2\times 10^{20} \, \rm cm^{-2}$.}
    \label{opt_beam.fig}
\vspace*{-0.25cm}\end{figure}

For each observed galaxy the WHISP pipeline reduction recipe produces
data cubes with an angular resolution of $\sim 12'' \times 12''/
\sin\delta$, $30'' \times 30''$ and $60'' \times 60''$.  The maps
with different resolutions are used to obtain some insight in the
effect of spatial resolution on the measurement of \hi\ cross
sectional areas (see section~\ref{resolution.sec}). The median 3-$\sigma$ 
column density sensitivity limit of the highest resolution \hi\ maps is $2.3 
\times 10^{20}~\icmsq$. At the lower
resolutions ($30''$ and $60''$) the column density sensitivities improve
with factors of approximately 3.5 and 10, respectively.
To illustrate the
resolution of our data, we calculate the total number of beams over
the optical area within the 25th magnitude isophote, and the number of
beams over the area with an \hi\ column density exceeding the DLA
limit. These statistics are presented in Fig.~\ref{opt_beam.fig}.
The median number of beams over the optical area is $\approx 40$,
whereas the median number of beams over the \hi\ area is $\approx
220$. We note that the total number of independent \hi\ column density
measurements above the DLA limit in our sample is $\approx 140,000$.


A possible concern about the selection of the WHISP sample is that the sample is based on the UGC, which implies that the parent sample from which the WHISP galaxies are chosen is optically selected. Since DLA systems are purely \hi-selected, our $z=0$ comparison sample of galaxies should ideally also be \hi-selected to guarantee that all possible optical selection biases are ruled out. However, \citet{zwaan2000} has shown that the properties of \hi-selected galaxies are not different from those of optically selected galaxies if compared at the same luminosity. Specifically, he showed that the gas richness, optical surface brightness, and scale length of optically selected and \hi-selected galaxies are indistinguishable if subsamples with similar median absolute magnitude are compared. Therefore, if the weighting of the sample is properly taken into account (using the luminosity function or \hi\ mass function) no strong biases should be introduced by using optically selected galaxies. As will be discussed in section~\ref{fn.sec}, we use morphological type-specific \hi\ mass functions to calculate weights for individual galaxies, which will ensure that the morphological type distribution of our weighted sample at a given \hi\ mass is similar to that of an \hi-selected galaxy sample. Small differences in the optical properties of optically and \hi-selected samples as a function of \hi\ mass might remain, but given the findings of \citet{zwaan2000} and the fact that we apply type-specific weighting, these differences are unlikely to bias are results significantly. This is also illustrated by the fact that the value for the redshift number density $dN/dz$ that we derive in section~\ref{dndz.sec} is consistent with what \citet{ryan-weber2003,ryan-weber2005} find based on a smaller sample of \hi-selected galaxies.

The field of view of the 21-cm WSRT observations is defined by the
diameter of the individual dishes and is approximately $30'$ across
(at half power beam width).  The WHISP data base contains many
instances of multiple \hi\ detections within one such field. We
separated these multiple detections into individual galaxies if both
\hi\ detections could be identified with optically identifiable
galaxies in the NASA/IPAC Extragalactic Database (NED). In most cases
the multiple detections are known pairs or groups of galaxies. If a
second \hi\ detection in the field targeted on a UGC galaxy could not
be associated with a catalogued galaxy, we did not separate this \hi\
detection but treated it as a companion \hi\ cloud to the known
optical galaxy and therefore included it in the total cross section
calculation, keeping information about the impact parameter at which
the \nhi\ occurs.

All 21-cm maps of the resulting sample of galaxies were transformed to
column density maps using the well-known equation
\begin{equation}
N_{\rm HI}=1.823\times 10^{18} \int T_{\rm B}(v) dv,
\end{equation}
where $T_{\rm B}$ is the observed brightness temperature. This
equation is valid for optically thin emission. For cold, high column
density ($>10^{21}\,\icmsq$) gas clouds, the optically thin
approximation might break down, which would result in slightly
underestimated \hi\ column densities. This effect is probably only
important for the very highest \hi\ column densities, in the range
where there are only very few DLA measurements. Therefore, it is
unlikely to affect our comparison between DLAs and local
galaxies. Furthermore, the measurement of total cross-sectional area
above the DLA column density limit is not affected at all by \hi\
optical depth effects.


\section{Searches for low redshift DLA galaxies}
The aim of this paper is to make a detailed comparison between the
local galaxy population and local DLA galaxies. The difficulty in this
comparison is that the number of identified DLA galaxies is very low,
at any redshift. There are now approximately 600 DLA systems known in
the Universe \citep[][]{prochaska2005}, only $\sim$ 50 of which are
at $z<1.7$. This number is especially small at low redshifts, because
the Ly-$\alpha$ line is only observable from the ground when it is
redshifted to $z>1.7$. Furthermore, due to the expansion of the
Universe, the expected number of absorbers along a line of sight
decrease with decreasing redshift, which naturally leads to a scarcity
of DLA systems at low $z$.

At high $z$ there is a substantial number of systems available, but
attempts to directly image their host galaxies have generally been
unsuccessful \citep[see
e.g.,][]{colbert2002,moller2002,prochaska2002}. A few positive
identifications do exist, mostly the result of HST imaging
\citep{fynbo1999,warren2001,kulkarni2000,moller2002}.  Possibly the
best data to date are available for three objects imaged with STIS by
\citet{moller2002}. They found emission from three DLA galaxies with
spectroscopic confirmation and concluded that the objects are
consistent with being drawn from the population of Lyman-break
galaxies.

Although the absolute number of DLAs at low $z$ is small, the success
rate for finding their host galaxies is better for obvious reasons:
the host galaxies are expected to be brighter and the separation on
the sky between the bright QSO and the DLA galaxy is likely
larger. The first systematic survey for low $z$ DLA host galaxies was
performed by \citet{lebrun1997}, who obtained HST $R$ and $B$-band
imaging of seven QSO fields with known DLAs. For most of their targets
they found likely galaxies, the properties of which span a large range
from LSB dwarfs to large spirals.  Similar searches for single systems
were done by \citet{burbidge1996} and \citet{petitjean1996}. The DLA
at $z=0.656$ in the sightline toward 3C336 was studied extensively by
\citet{steidel1997} and \citet{bouche2001}, but despite the deep HST
imaging in the optical and H$\alpha$, no galaxy has been identified.
All these studies lacked spectroscopic follow-up, so associations were
purely based on the proximity of identified galaxies to the QSO sight
line, which is not unique in many cases.  Optical and infrared imaging
combined with spectroscopic follow-up observations was first done by
\citet{turnshek2001}, who successfully identified two galaxies at
$z=0.091$ and $z=0.221$.  Using a similar technique, \citet{lacy2003}
also identified two low $z$ DLA galaxies.  \citet{chen2003}
demonstrated that it is possible to use photometric redshifts to
determine the association of optically identified galaxies with DLAs,
and applied this technique to six DLA systems. Finally,
\citet{rao2003} studied four DLA systems and were able to confirm two
DLA galaxies based on spectroscopic confirmation, one using a
photometric redshift, and one association was based on proximity to
the QSO.

Two attempts have been made to measure directly the cold gas contents
of low $z$ DLA systems by means of deep 21-cm emission line
observations \citep{kanekar2001, lane2000}. The problem with these
observations is that the 21-cm line is extremely weak so that with
present technology surveys are limited to the very local ($z<0.2$)
Universe. Both observations resulted in non-detections, which allowed
the authors to put upper limits to the \hi\ masses of approximately
$\mhi=2.3\times 10^9~\msol$, which is one third of the \hi\ mass of an
$L_*$ galaxy \citep{zwaan2003,zwaan2005}.

In addition to these surveys for low $z$ DLA galaxies, there are two
cases of high column density absorption systems found in local
galaxies.  \citet{miller1999} and \citet{bowen2001} study the
absorption line systems seen in NGC 4203 and SBS 1543+593, at
redshifts of $z=0.004$ and $z= 0.009$, respectively.  
For these systems, the
galaxy-QSO alignment was identified before the absorption was found
and in one case (NGC 4203) Ly-$\alpha$ absorption has not been seen to
date, but the structure and strength of observed metal lines resemble
those of DLAs.

All together, including the $z\approx 0$ systems NGC 4203 and SBS 1543+593, 
there are now 20 DLA
galaxies known at $z<1$. Table~\ref{props.tab} summarises the
properties of these galaxies. We have been fairly generous in
assembling this table because not all listed galaxies have been
spectroscopically confirmed. Also, some of the systems fall just below
the classical DLA column density limit of $\log \nhi=20.3$, but we
decided to include these because our 21-cm emission line data is also
sensitive below this limit. However, if in the remainder of this paper
when we refer to ``DLA column densities,'' we restrict our analysis to
$\log \nhi>20.3$ and disregard the identified systems with lower
column densities.  Some systems in our compilation have been studied
by several authors, in which case we choose the parameters from the
most recent reference.  The measurements of $L/L_*$ are mostly based
on $B$-band data, but in some cases these data were not available in
which cases we used $K$ or $R$-band. For the value of $L_*$ we adopted
the recent measurement of $M_B^*-5 \log h_{75}= -20.3$ from
\citet{norberg2002}, $M_R^*-5 \log h_{75}= -21.1$ from
\citet{lin1996}, and $M_K^*-5 \log h_{75}= -24.1$ from
\citet{cole2001}.  The morphological classifications are copied
directly from the relevant references and are very heterogeneous in
their degree of detail.  Therefore, we note that the $L/L_*$ values
and the listed types should be treated with caution and both
parameters have large uncertainties.  Our local DLA galaxy sample is
defined to have redshifts $z<1$, which is of course a random
choice. However, cutting the sample at lower redshift would result in
too small a sample to make a meaningful comparison with our nearby
galaxy statistics. The median redshift of the DLA galaxy sample is
$<z>=0.5$. In the analysis we ignore any evolutionary effect
in the redshift range $z=1$ to $z=0$. This assumption is justified by our finding in
section~\ref{dndz.sec}, that the cross section times comoving space
density of DLA systems is not evolving in this redshift
range. However, an important caveat is that this lack of evolution in
cross section puts no constraints on how gas is distributed in
individual systems, but only on the total gas content averaged over
the whole galaxy population. The mean properties of gas disks in
galaxies might evolve since $z=1$. Given the fact that the cosmic
star formation rate density drops by a factor $~8$ between $z=1$ and
$z=0$ \citep[e.g.,][]{hopkins2004}, it is possible that evolution of the
neutral gas disks actually takes place as well. To study the \hi\ mass
evolution of galaxies deep observations with future instrument such as
the Square Kilometer Array \citep[SKA, ][]{carilli2004} are
required. At present no observational constraints exist. Apart from
the evolution of the \hi\ disks, the optical properties of DLA host
galaxies could evolve between $z=1$ and $z=0$ as well. However,
\citet{chen2003} find that DLA galaxies do not show luminosity
evolution between $z=1$ and $z=0$, although the constraints are
limited by small number statistics.

Finally, one might be concerned about the process of choosing which
galaxy to attribute to a certain absorber \citep[see discussion
in][]{chen2003}. In some cases more than one candidate is available,
in which case the authors are forced to choose which one is the most
likely absorber.  The set of arguments on which this choice is based,
varies between different studies.  We note that this introduces some
randomness in our compilation presented in Table~\ref{props.tab}.

\begin{table*}
\centering
\begin{minipage}{130mm}
\caption{Properties of $z<1$ (sub-)DLA host galaxies.}
\label{props.tab}
\begin{tabular}{l l l c c l l l}
\hline
QSO & $z_{\rm abs}$ & log \nhi\ & $b$ & $L/L_*$ & Morphology & Reference & Redshift$^1$ \\
  &  & (cm$^{-2}$) & (kpc) & \\
\hline
        Ton 1480 & 0.004 & 20.34 &        7.9 & 0.44 &          early type &   a &   spectro-z\\ 
    HS 1543+5921 & 0.009 & 20.34 &        0.4 & 0.02 &          early type &   b &   spectro-z\\ 
       Q0738+313 & 0.091 & 21.18 &   $<$3.1 & 0.08 &                 LSB &   c &          no\\ 
    PKS 0439-433 & 0.101 & 20.00 &        6.8 & 1.00 &                disk & d,g,k &   spectro-z\\ 
       Q0738+313 & 0.221 & 20.90 &       17.3 & 0.17 &        dwarf spiral &  c,g &   spectro-z\\ 
    PKS 0952+179 & 0.239 & 21.32 &   $<$3.9 & 0.02 &           dwarf LSB &   e &     photo-z\\ 
    PKS 1127-145 & 0.313 & 21.71 &   $<$5.6 & 0.16 &  Patchy irr LSB$^2$ & e,g,l &   spectro-z\\ 
    PKS 1229-021 & 0.395 & 20.75 &        6.6 & 0.17 &             Irr LSB &   f &          no\\ 
       Q0809+483 & 0.437 & 20.80 &        8.3 & 2.80 &           giant Sbc & f,j,k &   spectro-z\\ 
     AO 0235+164 & 0.524 & 21.70 &       12.5 & 1.90 & late-type spiral $^2$ &  g,m &   spectro-z\\ 
     B2 0827+243 & 0.525 & 20.30 &       29.5 & 1.04 &    disturbed spiral &  e,g &   spectro-z\\ 
    PKS 1629+120 & 0.531 & 20.69 &       14.7 & 0.78 &              spiral &  e,g &   spectro-z\\ 
  LBQS 0058+0155 & 0.613 & 20.08 &        7.5 & 0.60 &              spiral &  h,k &   spectro-z\\ 
       Q1209+107 & 0.630 & 20.20 &        9.7 & 1.90 &              spiral &   f &          no\\ 
    HE 1122-1649 & 0.681 & 20.45 &       23.5 & 0.58 &             compact &   g &     photo-z\\ 
       Q1328+307 & 0.692 & 21.19 &        5.7 & 0.67 &                 LSB &   f &          no\\ 
 FBQS 1137+3907C & 0.720 & 21.10 &       10.3 & 0.20 &             spiral? &   i &   spectro-z\\ 
 FBQS 0051+0041A & 0.740 & 20.40 &       22.4 & 0.80 &             spiral? &   i &   spectro-z\\ 
     MC 1331+170 & 0.744 & 21.17 &       25.1 & 3.00 &      edge-on spiral &   f &          no\\ 
    PKS 0454+039 & 0.860 & 20.69 &        5.5 & 0.60 &             compact &  f,j &          no\\ 

\hline
\end{tabular}
\\{\small
References: (a) \citet{miller1999}; 
(b) \citet{bowen2001};
(c) \citet{turnshek2001};
(d) \citet{petitjean1996};
(e) \citet{rao2003};
(f) \citet{lebrun1997};
(g) \citet{chen2003};
(h) \citet{pettini2000};
(i) \citet{lacy2003};
(j) \citet{colbert2002};
(k) \citet{chen2005};
(l) \citet{lane1998};
(m) \citet{burbidge1996}\\
$^1$ Indication of whether the redshift of the DLA galaxy has been confirmed by either
spectroscopy or a photometric redshift measurement\\
$^2$ Probably arising in galaxy group}
\end{minipage}
\end{table*}

The conclusion from studying Table~\ref{props.tab} is that the sample
of low $z$ DLA galaxies spans a wide range in galaxy properties,
ranging from inconspicuous LSB dwarfs to giant spirals and even early
type galaxies. Obviously, it is not just the luminous, high surface
brightness spiral galaxies that contribute to the \hi\ cross section
above the DLA threshold, although it is these 
galaxies that contribute most to the total comoving
density \ohi.


\section{Calculating the column density distribution function - \fnhi} \label{fn.sec}
In QSO absorption line studies, the column density distribution \fnhi\
  function is defined such that $f(\nhi) d\nhi dX$ is the number of
  absorbers with column density between $\nhi$ and $\nhi+d\nhi$ over
  an absorption distance interval $dX$. The analysis in this paper
  derives \fnhi\ from the statistics of \hi\ emission line
  distributions in the local galaxy population in order to predict the
  effectiveness of present day galaxies at accounting for the
  absorption line statistics.  The parameter $dX$ assures that
  $f(\nhi)$ counts the number of absorbers per co-moving unit of
  length, but since $dX/dz=1$ at $z=0$, we can replace $dX$ with $dz$
  in our calculation of $f(\nhi)$.  The local $f(\nhi)$ can now be
  calculated from
\begin{equation}
f(\nhi)=\frac{c}{H_0} \frac{\sum_i  \Phi({\bf x}_i) w({\bf x}_i) A_i (\log \nhi)}{\nhi \, \ln 10 \,\Delta \log \nhi  }.
\end{equation}
Here, $\Phi({\bf x}_i)$ is the space density of objects with property 
${\bf x}_i$ equal to that of galaxy $i$.  In reality, the parameter 
${\bf x}$ could be the \hi\ mass
or the optical luminosity of the galaxies in the sample, so that
$\Phi({\bf x})$ is the \hi\ mass function or the optical luminosity
function of galaxies in the local Universe.
$A_i(\log \nhi$) is the area function that describes for
galaxy $i$ the area in $\rm Mpc^{2}$
corresponding to a column density in the range $\log \nhi$ to $\log \nhi+\Delta \log \nhi$.
In practice, this is simply calculated by summing for each galaxy 
the number of pixels in a certain $\log \nhi$ range multiplied by the physical
area of a pixel. The function $w({\bf x}_i)$ is a weighting function that takes into account the varying number of galaxies across the full stretch of $\bf x$, and is calculated by taking the reciprocal of the number of galaxies in the range $\log {\bf x}-\Delta/2$ to  
$\log {\bf x}+\Delta/2$, where $\Delta$ is taken to be 0.3. The results are very
insensitive to the exact value of $\Delta$.
The summation  sign denotes
a summation over all galaxies in the
sample. Finally, $c/H_0$ converts the number of systems per Mpc to
that per unit redshift. Note that dependencies on $H_0$ disappear in
the final evaluation of $f(\nhi)$.

The WHISP sample that is used in this analysis is somewhat biased toward early type galaxies (S0 to Sb) because these were specifically targeted for the projects described in \citet{vanderhulst2001}. If we were to apply a simple weighting scheme using a luminosity or \hi\ mass function, early type galaxies would be given too much weight in the calculation of the $f(\nhi)$. To avoid this problem, we choose to apply type-specific \hi\ mass functions as published by \citet{zwaan2003}. Morphological types are available in LEDA for all but 18 galaxies in the WHISP sample. We divide the WHISP sample into five subsets of galaxies: E-S0, Sa-Sb, Sbc-Sc, Scd-Sd, and Sm-Irr. Visual inspection of the unclassified galaxies learned that these were mostly dwarf irregular or small compact galaxies and hence we put these in the Sm/Irr subset. We note that the difference between the $f(\nhi)$ calculated using the weights based on one \hi\ mass function for all WHISP galaxies and that based on type-specific weighting is very small.

\begin{figure}\centering
\includegraphics[width=1.00\columnwidth,trim=0cm 1.0cm 0.cm 0.0cm ]{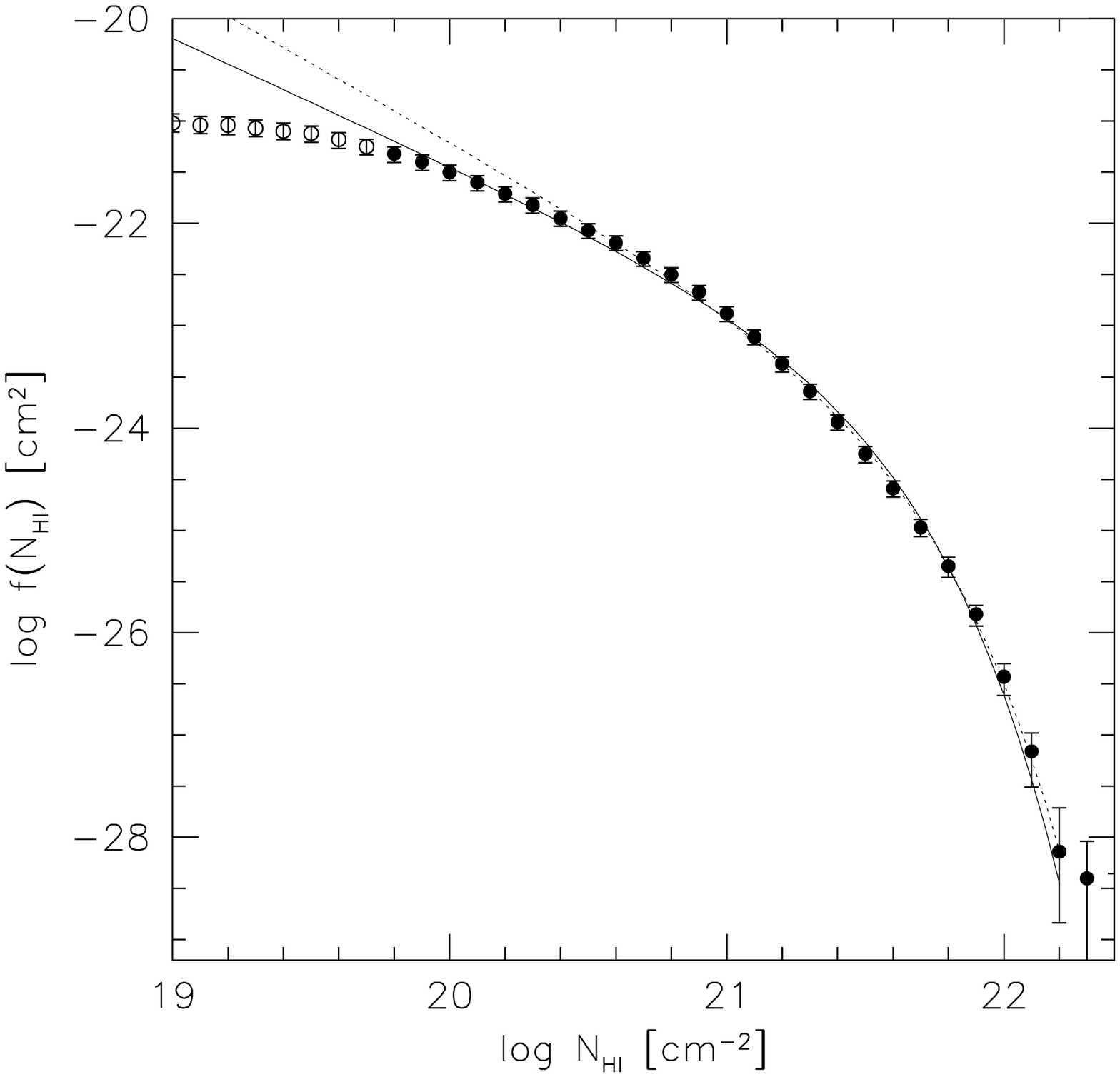}
    \caption{The \hi\ column density distribution function at $z=0$
    from 21-cm emission line observations. Error bars include
    uncertainties in the \hi\ mass function as well as counting
    statistics and indicate $1\sigma$ uncertainties.  Open symbols
    correspond to measurements below our sensitivity limit. The solid
    line is a gamma distribution fit (see Eq.~\ref{fnfit.eq}) to all
    points above $\log \nhi>19.8$, the dotted line is the same but
    restricted to $\log \nhi>20.3$. }
    \label{fn_fit.fig}
\vspace{-0.25cm}\end{figure}

Fig.~\ref{fn_fit.fig} shows our measured column density distribution
function $f(\nhi)$. We use a binning of $\log \nhi=0.1\, \rm dex$. The
error bars indicate $1\sigma$ uncertainties and include uncertainties
in the \hi\ mass function normalisation from \citet{zwaan2003} as well
as counting statistics of the WHISP sample. The open symbols indicate
column densities below $\log\nhi=19.8$, which we adopt as our
sensitivity limit (see Section~\ref{resolution.sec} for a discussion
on this limit).  Following \citet{pei1995},
\citet{storrie-lombardi1996}, \citet{storrie-lombardi2000}, and
\citet{peroux2003}, we fit $f(\nhi)$ with a gamma distribution,
analogous to the Schechter function, which is used for fitting luminosity functions and \hi\ mass functions:
\begin{equation}
f(\nhi)=(f^*/\nhis)(\nhi/\nhis)^{-\beta}e^{-\nhi/\nhis}.
\label{fnfit.eq}\end{equation}
This gives an excellent fit to our data with parameters:
$\beta=1.24$, $\log\nhis=21.2 \,\icmsq$, and $f^*=  0.0193$ if we use all
data points above $\log \nhi>19.8$. The parametric $f(\nhi)$ is shown
as a solid line in Fig.~\ref{fn_fit.fig}. If we restrict the fit to
column densities above the DLA limit, we find $\beta=1.52$,
$\log\nhis=21.3 \,\icmsq$, and $f^*=0.0137$, which is shown by the dotted
line.  We note that there is no physical motivation for fitting the
$f(\nhi)$ with a gamma distribution other than that it provides a
reasonable fit. Neither is there a physical reason to fit $f(\nhi)$
with a power law, which is traditionally done
\citep[e.g.,][]{tytler1987}. However, the good gamma distribution fit
demonstrates that at high \hi\ column densities ($\log \nhi>21$), the
$f(\nhi)$ deviates strongly from a traditional single power-law
fit. This is consistent with the form of the cut-off identified by
\citet{peroux2003} based on a large sample of DLAs and sub-DLAs.

The total \hi\ mass density can be calculated by integrating over the
column density distribution function. When using the parametrised
gamma distribution, it can be easily seen that the \hi\ mass density
can be calculated as
$$
\rho_{\rm HI}=\Gamma(2-\beta)f^* \nhis\, m_{\rm H}  \, H_0/c,
$$ where $m_{\rm H}$ is the mass of the hydrogen atom and $\Gamma$ is
the Euler gamma-function. Using the best-fit parameters of our gamma distribution fit
to $\log \nhi>19.8$, we find $\rho_{\rm HI}=6.8\times 10^7 \msol \rm
Mpc^{-3}$, in excellent agreement with \citet{zwaan2003}. This may not
come as a surprise, since we used the \hi\ mass functions from that
paper to calibrate \fnhi. Nonetheless, the result of this calculation
is a good consistency check and assures that the normalisation of our
$f(\nhi)$ is correct \citep[see also][]{rao1993}.

There are a few interesting features in Fig.~\ref{fn_fit.fig}. First,
our measured points begin to deviate from the power law end of the
gamma distribution fit at column densities $\log \nhi < 19.8$. The first
obvious explanation for this is that in this range of \nhi\ we start
to loose sensitivity in the 21-cm maps. There might also be a more
physical origin of this effect, which is the expected ionisation of
the outer \hi\ disks of galaxies by the metagalactic UV-background
below $\log \nhi \approx 19.5$
\citep{corbelli1993,maloney1993}. \citet{corbelli2002} argue that if
\fnhi\ is corrected to include \hii, it would follow a power-law down
to $\nhi=10^{17}~\icmsq$.  Our 21-cm maps lack the sensitivity to
study this effect in detail. At the high \nhi\ end in
Fig.~\ref{fn_fit.fig}, the \fnhi\ drops off exponentially, which is
faster than $\nhi^{-3}$ above $\log \nhi=21.6$. A fall-off of
$\nhi^{-3}$ is theoretically expected for randomly oriented gas disk,
irrespective of their radial \nhi\ distribution
\citep[see][]{milgrom1988, fall1993, zwaan1999}. Three independent effects can
cause deviation of the $\nhi^{-3}$ expectation: 1) beam smearing in
the 21-cm maps can smooth out the very highest column densities; 2)
galaxies are not infinitely thin (this is assumed in the \fnhi\
calculation of randomly oriented disks); and 3) the \hi\ gas is not
optically thin at the highest column densities. In the remainder of
this paper, our possible biases at the extreme edges of $f(\nhi)$ are
not important because we only compare our data with the absorption
line data in the intermediate \nhi\ range. Furthermore, the
cosmological \hi\ mass density is also dominated by the intermediate
\nhi\ values, as will become clear in section~\ref{fn_hiz.sec}.

\subsection{The effect of spatial resolution on \fnhi}{\label{resolution.sec}}
One concern when comparing column densities measured from 21-cm maps
with those measured from absorption line systems is the large
difference in spatial resolution.  The median distance of the WHISP
sample is 20 Mpc, implying that for the highest resolution maps the
synthesised beam corresponds to $\approx 1.3$ kpc.  This is almost
$10^5$ times greater than the diameter of the optical emission region
of a QSO ($\leq 0.03$ pc, e.g. \citealt{wyithe2002}), the area probed
by DLA column density measurements.

Despite this disparity in resolution, 21-cm emission and
Lyman $\alpha$ absorption measurements generally seem to be in
agreement. \citet{dickey1990} noted a consistency between
Lyman $\alpha$ absorption measurements towards high latitude stars and
21-cm emission measurement in the Galaxy. There is only one example
where a column density from both 21-cm emission and DLA absorption has
been measured of the same source. The DLA galaxy SBS 1543+593 has
Lyman $\alpha$ absorption measured at $2.2\times10^{20}\, \icmsq$
\citep{bowen2001}, whereas the 21-cm emission at the position of the
background QSO is measured to be $5\times10^{20}\, \icmsq$
\citep{chengalur2002}.

An alternative method to test the effect of resolution is to spatially
smooth high resolution 21-cm maps and measure the resulting column
densities and \fnhi.  \citet{ryan-weber2005b} used a high resolution
21-cm data cube of the Large Magellanic Cloud and convolved it with
circular 2-dimensional Gaussians of various widths in the spatial
plane. The $\nhi$ distribution was then re-measured and \fnhi\ was
re-calculated, based on the new, low resolution column
densities. Lowering the resolution was found to have a truncating
effect on the \nhi\ distribution, generally decreasing the occurrence
of the highest column densities.

\begin{figure}\centering
\includegraphics[width=1.00\columnwidth,trim=0cm 1.0cm 0.cm 0.0cm ]{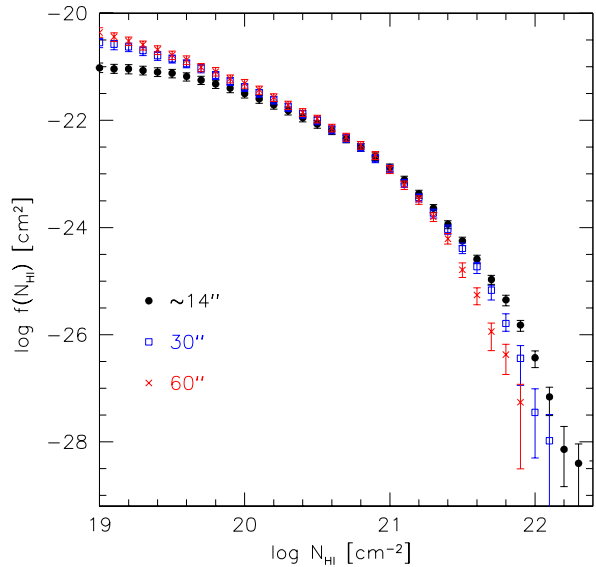}
    \caption{The effect of spatial resolution of the 21-cm line
    observations on the \hi\ column density distribution function at
    $z=0$. Filled circles correspond to the highest resolution, which
    is approximately $14''$, open squares and crosses are for the
    spatially smoothed data, to resolutions of $30''$ and $60''$,
    respectively.}
        \label{comp_res.fig}
\vspace{-0.25cm}\end{figure}

For our set of WSRT maps, we can test the effect of lowering the
spatial resolution.  This is shown in Fig~\ref{comp_res.fig}, where we
present the \fnhi\ distribution derived from the $30''$ and $60''$ maps,
along with that from the original resolution maps.  It can be seen
that going to lower resolution, the highest column densities are
smoothed away and this flux appears again as lower column
densities. Effectively, this causes a `tilt' of \fnhi. From this
sample it is impossible to estimate if this tilting would continue if
even higher resolution maps would be used. However, for our comparison
with DLA data the effect of resolution is probably unimportant because
the effects are minimal in the intermediate \nhi\ range where we make
the comparison.

The redshift number density, $dN/dz$ (see section \ref{dndz.sec}) is
also only minimally affected by resolution. The LMC study by
\citet{ryan-weber2005b} showed that $dN/dz$ (at a limit of $\log
\nhi=20.3$) increases on the order of 10 per cent when the spatial
resolution is changed from 15 pc to 1.3 kpc, the typical resolution of
the WHISP sample.  A 10 per cent change in the measured $dN/dz$ for
the WHISP sample is well within the errors quoted in section
\ref{dndz.sec}.

Finally, we comment on our adopted sensitivity limit of $\log \nhi=19.8$
for the column density distribution function. In Section~\ref{whisp.sec} we 
quoted a formal 3$\sigma$ sensitivity limit of $\log \nhi=20.3$ for the high 
resolution maps, which implies 
that below this limit we might start to underestimate \fnhi. However, 
Fig.~\ref{comp_res.fig} shows that \fnhi\ based on lower resolution data
is marginally higher in the range $19.8<\log\nhi<20.3$, and only
significantly deviates from the high resolution curve for $\log \nhi<19.8$. 
As explained above, the higher \fnhi\ values for lower resolution data
are not only the result of these data having lower sensitivity limits and thus
picking up lower column density \hi, but also because small regions of high \nhi\
gas are smoothed away and appear again as lower column densities. This also
explains why small differences between \fnhi\ based on the different resolution
data already become apparent above $\log \nhi=20.3$.
Therefore, if we were to construct a ``hybrid''  \fnhi\ using high \nhi\ results from high 
resolution  data and low \nhi\ from low resolution data, we would double-count some of 
the \hi\ atoms. We choose to simply adopt a sensitivity limit of $\log\nhi=19.8$,
and note that the low end ($\log\nhi<20.3$) of \fnhi\ might be slightly underestimated, but
by not more  than 20 per cent. This does not influence any of our conclusions 
presented in this paper.


\section{Comparison of \fnhi\ with high \redshift\ data} \label{fn_hiz.sec}
\begin{figure}\centering
\includegraphics[width=1.00\columnwidth,trim=0cm 1.0cm 0.cm 0.0cm ]{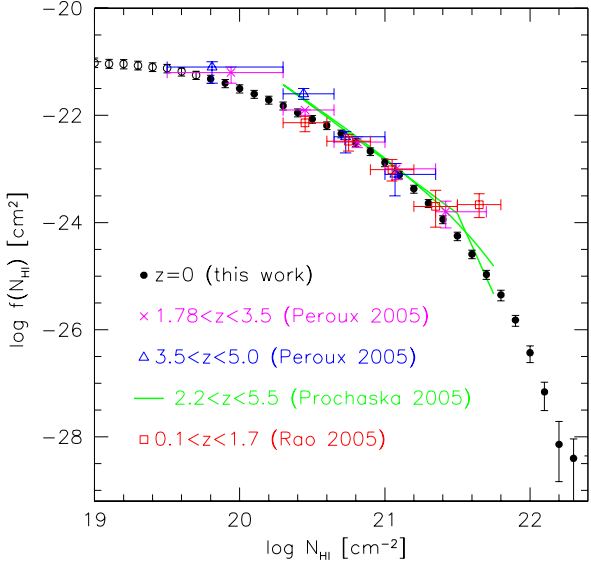}
    \caption{Comparison between the \hi\ column density distribution
    function at $z=0$ and higher redshifts.  The solid points are from
    our 21-cm emission line observations and are the same as in Fig.
    \ref{fn_fit.fig}.  The higher redshift points are from
    \citet{peroux2005} and \citet{rao2005}.  Horizontal errorbars
    indicate the bin sizes.  The curves show the gamma distribution and double
    power-law fits to the SDSS-DR3 results from \citet{prochaska2005}
    and cover a redshift range of $2.2<z<5.5$. All results based on a
    $\Omega_{\rm m}= 0.3$ and $\Omega_{\Lambda}=0.7$ cosmology.  }
    \label{fn_peroux.fig}
\vspace*{-0.25cm}\end{figure}

In Fig.~\ref{fn_peroux.fig} we reproduce the \fnhi\ measurements from
our analysis of 21-cm maps and compare these with data at higher
redshifts from \citet{peroux2005}, \citet{rao2005} and
\citet{prochaska2005}. The \citet{peroux2005} measurements of \fnhi\
below the DLA limit are the result of their new UVES survey for
``sub-DLAs''. We choose here to only plot the points for $\log
\nhi>19.8$, which roughly corresponds to our sensitivity
cut-off. Their higher column density points stem from the combined
data of \citet{storrie-lombardi1996}, \citet{storrie-lombardi2000},
and \citet{peroux2001}, together based on $\approx 100$ $z>4$ quasars.
The \citet{prochaska2005} data are from an automatic search for
$z>2.2$ DLA systems in SDSS-DR3 and also include the results from
previous DLA surveys.  We represent their results by both a gamma distribution
fit and a double power-law fit to the whole data set in the redshift
range $2.2<z<5.5$.  The intermediate redshift points from
\citet{rao2005} are based on \mgii-selected DLA systems.  All
calculations are based on a $\Omega_{\rm m}= 0.3$,
$\Omega_{\Lambda}=0.7$ cosmology. The surprising result from this
figure is that there appears to be only {\em very mild evolution in
the intersection cross section of \hi\ from redshift $z\sim 5$ to the
present}. This is a conclusion in stark contrast to earlier works that
claimed strong evolution in the DLA cross section
\citep[e.g.,][]{wolfe1986,lanzetta1991}. We will come back to this
weak evolution in section~\ref{dndz.sec} in which we discuss the
redshift number density.

\begin{figure}\centering
\includegraphics[width=1.00\columnwidth,trim=0cm 1.0cm 0.cm 0.0cm ]{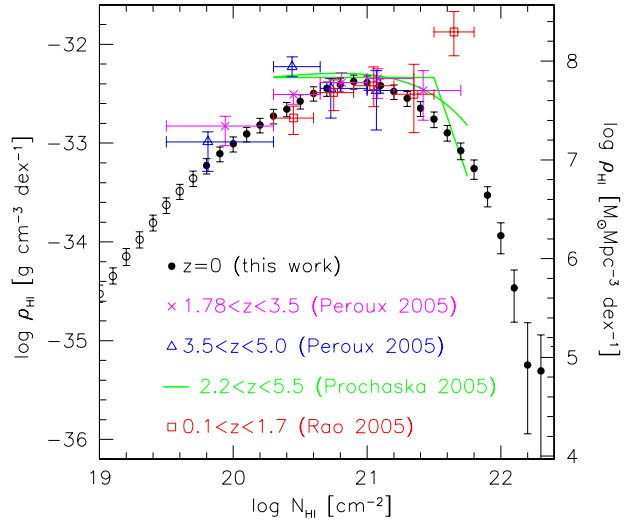}
    \caption{The \hi\ mass density contained in systems of different
    column density per decade of \nhi. Solid circles are for $z=0$ and
    are from our data, other symbols are for higher redshifts and are
    derived from \citet{peroux2005} and \citet{rao2005}. The curves
    are the converted fits to the SDDS-DR3 results from
    \citet{prochaska2005}.}
    \label{fn_mass.fig}
\vspace*{-0.25cm}\end{figure}

In Fig.~\ref{fn_mass.fig} we plot the contribution of systems with
different column densities to the integral \hi\ mass density
$\rho_{\rm HI}$. At $z=0$ it is clear that systems with column
densities $\nhi \sim 10^{21}\, \icmsq$ dominate the \hi\ mass
density. At higher redshifts ($z>1.7$), the uncertainties are much
larger, but also there it appears that $\nhi \sim 10^{21}\, \icmsq$
systems contribute most, although the \citet{prochaska2005} data
suggest that column densities between $\log \nhi=20.3$ and $21.3$
contribute almost evenly.  Any possible differences between the
results at various redshifts are much more pronounced in
Fig.~\ref{fn_mass.fig} than in Fig.~\ref{fn_peroux.fig} because
the vertical scale is stretched from nine decades to only four
decades. The one point that clearly deviates is the highest \nhi\
point from \citet{rao2005} at $\log\nhi=21.65$.  This elevated
interception rate of high \nhi\ \mgii-selected intermediate redshift
DLAs is present in both redshift bins that \citet{rao2005} distinguish
($0.1<z<0.9$ and $0.9<z<1.7$), and is also reported in their earlier
work \citep{rao2000}.  Fig.~\ref{fn_mass.fig} very clearly
demonstrates that this point dominates the \ohi\ measurement at
intermediate redshifts. It is therefore important to understand
whether the \mgii-based results really indicate that high column
densities ($\log\nhi\sim21.65$) are rare at high redshift, then indeed
become more abundant at intermediate redshifts ($0.1<z<1.7$) and
subsequently evolve to the scarce numbers at $z=0$. Alternatively, the
high \fnhi\ point might be a result of yet unidentified selection
effects introduced by the \mgii-selection, although presently there
are no indications that support this \citep{peroux2004,rao2005}.

\begin{figure}\centering
\includegraphics[width=1.00\columnwidth,trim=0cm 1.0cm 0.0cm 0.5cm ]{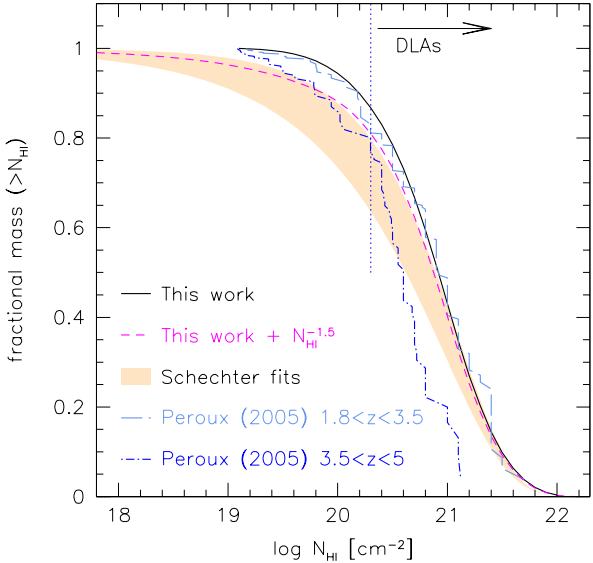}
    \caption{Cumulative \hi\ mass distribution in systems with column
    density $> \nhi$.  The solid line is based on our data alone, the
    dashed line includes a correction for \hi\ gas with column
    densities below our sensitivity limit and indicates that $\approx
    81$ per cent of the integral mass in \hi\ is in classical damped systems
    with $\log \nhi>20.3\, \rm cm^{-2} $.  The shaded area indicates
    the distribution based on the gamma distribution fits to our \fnhi\
    (Figure~\ref{fn_fit.fig}), where the lower boundary corresponds to
    the fit to $\log\nhi>20.3$ and the upper boundary to the fit to
    $\log\nhi>19.8$.  The long-dashed and dash-dotted lines are from
    \citet{peroux2005} and are for redshift ranges $1.8<z<3.5$ and
    $z>3.5$, respectively.  }
    \label{mass_N.fig}
\vspace*{-0.25cm}\end{figure}

To further investigate the relative contribution of different column
densities to the \hi\ mass density, we plot in Fig.~\ref{mass_N.fig} the
cumulative \hi\ mass density as a function of column density.  The solid
curve is based on our measured \fnhi\ points, not on the gamma
distribution fit to the data. From this curve alone, we would conclude
that at $z=0$ the fractional mass in systems with $\log \nhi > 20.3$
(classical DLAs) is 86 per cent. However, the contribution of gas with column
densities below our sensitivity limit of approximately $\log
\nhi=19.8$ is very uncertain.  Unfortunately, only selective regions have been imaged to low column density limits, but no large scale 21-cm emission line surveys that reach sensitivities below $\log \nhi=19.8$ are available yet. For example, the M31 environment has been imaged with the GBT and the WSRT at resolutions between 50 pc and 11 kpc \citep[][and references therein]{braun2004}, covering the column density range $17  <\log \nhi < 22$. However, the \hi\ emission in this region is mostly due to discrete High Velocity Clouds (HVCs) physically associated to M31. These surveys may not be representative of all low column density \hi\ gas at $z=0$, and do not provide good constraints on the shape of \fnhi. To obtain a rough estimate of the contribution of gas below our sensitivity limit, we extrapolate our measured \fnhi\ below $\log\nhi=19.8$ with $\fnhi \propto \nhi^{-1.5}$ \citep[per e.g., ][]{tytler1987} and find that the fractional mass in systems with $\log \nhi  >20.3$ is $\approx 81$ per cent, and hence only $\approx 19$ per cent of the \hi\ atoms in the local UniverUniverse are in sub-DLA column densities. For illustrative purposes, we also show as a shaded area the cumulative distribution functions based on the gamma distribution fits to \fnhi\ as presented in Figure~\ref{fn_fit.fig}. Here, the fit to $\log\nhi >19.8$ corresponds to the upper boundary and the fit to $\log\nhi>20.3$ to the lower boundary to this area. Obviously, simply extrapolating the gamma distribution fits to \fnhi\ below the DLA limit results in a severe overestimation of the number of \hi\ atoms in sub-DLA systems.

The higher redshift curves are based on the data of
\citet{peroux2005}.  With their new column density measurements of
individual sub-DLAs, these authors improved the earlier estimates of
\fnhi\ presented in \citet{peroux2003} and now show that the
fractional contribution of sub-DLAs to \ohi\ is approximately 20 per cent at
all redshifts $z>1.8$.  These curves also indicate that at high
redshifts the highest column densities are relatively rarer, as can
also be seen in Figure~\ref{fn_peroux.fig}.  \citet{prochaska2005}
estimate the fractional contribution of sub-DLA systems to the
integral mass density by combining their measured \fnhi\ with the
redshift number density of sub-DLAs from \citet{peroux2001}. They
conclude that the sub-DLAs (or super-LLSs in their terminology)
contribute between 20 and 50 per cent over the redshift range 2.2 to 5.5, but
emphasise that their single power-law fit to \fnhi\ might even
underestimate this fraction.  Obviously, the importance of sub-DLA
systems at high redshift remains a topic needing further
clarification in the future.

The fact that from our high resolution 21-cm maps we find that
$\approx 81$ per cent of the \hi\ mass density at $z=0$ is in column
densities above the DLA limit, implies that \ohi\ measurements from
blind 21-cm surveys overestimate $\Omega_{\rm DLA}(z=0)$ by only
$\approx 23$ per cent.  When comparing measurements of the atomic gas mass
density at high and low redshifts, it is important that the $z=0$
value be corrected with a factor 0.81 when compared to measurements
from DLAs. On the other hand, when at higher redshift the sub-DLA
contribution is taken into account \citep[e.g.,][]{peroux2005}, no
correction of the $z=0$ value is required. We refer to
\citet{prochaska2005} for various definitions of $\Omega_{\rm gas}$
and their recommended usage.

%
\section{The redshift number density at $\redshift=0$}\label{dndz.sec}
\begin{figure}\centering
\includegraphics[width=1.00\columnwidth,trim=0cm 1.0cm 0.0cm 0.5cm ]{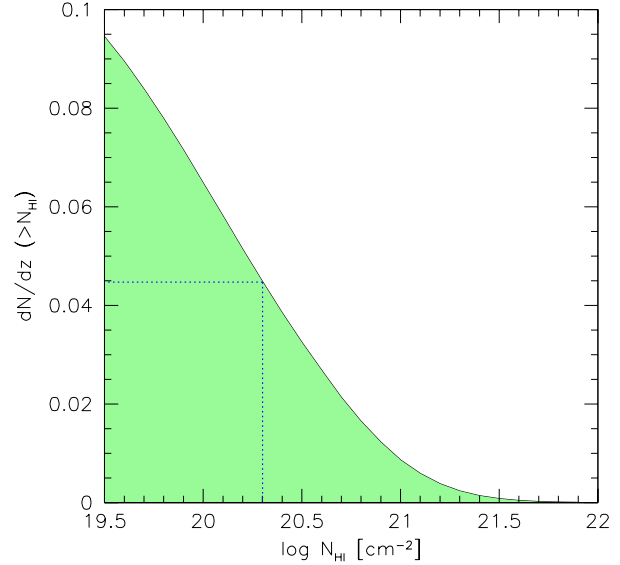}
    \caption{Integral redshift number density $dN/dz$ as a function of
    \hi\ column density cut-off.  For $\log \nhi>20.3 \, \rm cm^{-2}$
    we find $dN/dz=0.045$, dropping rapidly for higher column
    densities.}
    \label{dndz_N.fig}
\vspace*{-0.25cm}\end{figure}

The redshift number density $dN/dz$ for column densities larger than
$\nhi$ can be calculated from the $z=0$ \fnhi\ distribution by summing
over all column densities larger than $\nhi$:
\begin{eqnarray}
dN/dz  &=& \int_{N'_{\rm HI}>N_{\rm HI}} f(N'_{\rm HI}) \, d{N'_{\rm HI}}\\
&=&\frac{c}{H_0} \sum_{N'_{\rm HI}>N_{\rm HI}}
\frac{\sum_i  \Phi({\bf x}_i) w({\bf x}_i) A_i (\log N'_{\rm HI})}{\Delta \log N'_{\rm HI}  },
\end{eqnarray}
where definitions of $\Phi$, $w$, and $A_i$ are equal to those in section~\ref{fn.sec}.
In Fig.~\ref{dndz_N.fig} we show $dN/dz$ as a function of \nhi. For
column densities in excess of the DLA limit of $\log \nhi=20.3$ we
find $dN/dz=0.045\pm 0.006$, where the $1\sigma$ uncertainty is
dominated by the uncertainty in the parameters of the \hi\ mass
function that is used to calibrate \fnhi. For larger column densities
$dN/dz$ drops rapidly, systems with $\log \nhi>21$ contribute only
20 per cent to the total DLA cross section.  Our measurement of $dN/dz $
translates into a mean cross section density of $<n\sigma>=(1.13 \pm
0.15)\times 10^{-5} \,\rm Mpc^{-1}$, or to a mean free path between
absorbers of $l=<n\sigma>^{-1}=88 \pm 12 \,\rm Gpc$.

The value of $dN/dz (z=0)=0.045$ agrees very well with our previous
measurement of $dN/dz=0.042\pm 0.015$, based on a much smaller sample
of galaxies \citep{zwaan2002}.  \citet{rosenberg2003} used
\hi-selected galaxies to find $dN/dz=0.053\pm 0.013$, also in good
agreement with our value. Their slightly higher value is probably the
result of the steeper \hi\ mass function slope that these authors use
to calibrate their \fnhi.  \citet{ryan-weber2003} also used a sample
of \hi-selected galaxies observed with the ATCA and applied the same
HI mass function normalisation as in the present analysis, to find
$dN/dz=0.046^{+0.03}_{-0.02}$ \citep[see also][]{ryan-weber2005}.  An
earlier calculation by \citet{rao1993} resulted in a much lower value
of $dN/dz=0.015$. This discrepancy arises partly because these authors
limited their analysis to large, optically bright galaxies, whereas
the newer values take into account dwarf and low surface brightness
(LSB) galaxies, and partly because their calculation is based on the
Gaussian luminosity function from \citet{tammann1985}, which has a
lower normalisation than the most recent \hi\ mass function
measurements.  

It has been noted before that the area of the \hi\ disk of galaxies
above the DLA limit correlates very tightly to the \hi\ mass
\citep[e.g.,][]{broeils1992,rosenberg2003}. We can make use of this
correlation to obtain a alternative measurement of $dN/dz$. Suppose
that the projected \hi\ area $A$ correlates with \mhi\ as $A\propto
\beta \mhi$, and $A^*$ is the projected area of an $\mhi^*$ galaxy,
then
\begin{eqnarray}
dN/dz &=& \frac{c}{H_0} \int \theta(\mhi) A(\mhi) d\mhi \\
            &=& \frac{c}{H_0} \,A^* \theta^* \Gamma(1+\alpha+\beta),
\end{eqnarray}
where $\theta*$ and $\alpha$ are the normalisation and the low-mass
power-law slope of the \hi\ mass function, respectively. For our WHISP
sample we find that $A^*=992\, \rm kpc^{2}$. Using this value, the
\citet{zwaan2003} \hi\ mass function parameters ($\alpha=-1.30$ and $\theta^*=0.0086 \,{\rm Mpc}^{-3}$), and $\beta=1$, we find that
$dN/dz=0.044$. Quite surprisingly, this simplistic approach yields a
result very close to our measurement of $dN/dz =0.045$ based on a much
more detailed analysis.

\subsection{Evolution of $dN/dz$}
\begin{figure}\centering
\includegraphics[width=1.00\columnwidth,trim=0cm 0.2cm 0.0cm 0.5cm  ]{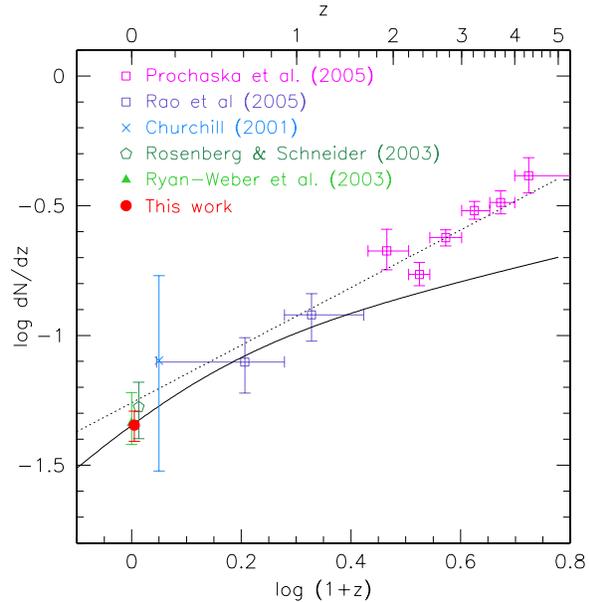}
    \caption{Number density of DLAs per unit redshift as function of
    redshift. All points at $z>0.1$ are from UV and optical surveys
    for DLA systems or MgII systems (see text). The points at $z\sim
    0$ are from techniques similar to that presented in this work. The
    dashed line shows the best fit to $dN/dz$ from
    \citet{storrie-lombardi2000}. The solid line represents ``no
    evolution in the product of cross section and co-moving space
    density'' for a cosmology with $\Omega_{\rm m}= 0.3$ and
    $\Omega_{\Lambda}=0.7$. This line is scaled vertically so as to
    fit our $z=0$ point.}
    \label{dndz.fig}
\vspace*{-0.25cm}\end{figure}

We now have a robust measurement of the redshift number density at
$z=0$. How does this value compare to DLA measurements at higher
redshifts? In Fig.~\ref{dndz.fig} we show the combined results of high
and low redshift $dN/dz$ measurements from different surveys. At
$z=0$, we plot our points as well as those from
\citet{ryan-weber2003,ryan-weber2005} and \citet{rosenberg2003}. All
these local values are based on analyses of \hi\ 21-cm maps of
galaxies, have small error bars and are consistent with each other.
The points at $z\approx 0.6$ and $z\approx 1.2$ are from
\citet{rao2005} based on their $z<1.65$ HST survey of \mgii\ selected
systems from the SDSS-EDR. Their total sample of $z<1.65$ DLAs is now
41, which is a considerable improvement over the original sample of 16
DLAs presented in \citet{rao2000}.  The underlying assumption that the
incidence of DLA systems can be derived from \mgii\ surveys relies on
the empirical fact that all DLA systems show \mgii\ absorption,
whereas the reverse is not true. The known incidence of \mgii\ systems
can therefore be used to bootstrap the DLA statistics. A similar
technique was used by \citet{churchill2001} , who found
$dN/dz=0.08^{+0.09}_{-0.05}$ at a median redshift of $z=0.05$.  This
number is based on HST data on four \mgii\ absorbers.  The high
redshift points from \citet{prochaska2005} are the result of an
automatic search for $z>2.2$ DLA systems in SDSS-DR3 and also include
the results from previous DLA surveys as summarised in
\citet{storrie-lombardi2000} and \citet{peroux2003}.

Also shown in Fig.~\ref{dndz.fig} by a dashed line is the fit
$dN/dz=0.055 (1+z)^{1.11}$ from \citet{storrie-lombardi2000}, which
agrees reasonably well with the $z=0$ value. It is often quoted in the
literature that this fit indicates ``no intrinsic evolution in the
product of space density and cross-section'' of damped absorbers,
which is another way of saying that the number of systems per
co-moving unit of length does not evolve.  This statement is based on
the fact that for a $q_0=0$ Universe, $dX/dz =1+z \approx
(1+z)^{1.11}$. However, for a modern non-zero $\Lambda$ Universe,
$dX/dz$ is given by
\begin{equation}
\frac{dX}{dz}=\frac{(1+z)^2}{\sqrt{\Omega_{\rm M}(1+z)^3-(\Omega_{\rm M}+\Omega_{\Lambda}-1)(1+z)^2+\Omega_{\Lambda}}},
\end{equation}
which starts to deviate significantly from the $q_0=0$ prediction for
$z>1$.  For no evolution in the number of systems per co-moving unit
of length, we expect
\begin{equation}
\frac{dN}{dz}=\left ( \frac{dN}{dz} \right ) _{z=0} \frac{dX}{dz}.
\end{equation}
The solid line in Fig.~\ref{dndz.fig} represents this run of $dN/dz$
as function of $z$ for an $\Omega_{\rm m}= 0.3$ and
$\Omega_{\Lambda}=0.7$ Universe. We normalise this line to our
$dN/dz(z=0)=0.045$ measurement.  With respect to this $\Omega_{\rm m}=
0.3$, $\Omega_{\Lambda}=0.7$ prediction, there is only weak evolution
in the comoving incidence rate from $z\sim 4$ to the present
time. Between $z\sim 1.5$ and $z=0$ there is no evidence for evolution at
all, between $z\sim 4$ and $z=0$, the evolution in the comoving
incidence rate is approximately a factor 2.

This conclusion contrasts with previous claims that the local galaxy
population cannot explain the DLA incidence rate.  For example,
estimates of the cross section to DLA absorption in local galaxy disks
by \citet{wolfe1986} and \citet{lanzetta1991} indicated that there
should be evolution of at least a factor of 2 to 4 (depending on the
value of $q_0$). Likewise, the analysis of \citet{rao1993} pointed
toward strong evolution in the incidence rate since $z=2.5$.  Part of
the reason for our conclusion being different from older works, is the
change in cosmological parameters. For no evolution in the comoving
number density in a $q_0=0.5$ cosmology, the allowed change in $dN/dz$
is much smaller than that for a modern $\Omega_{\rm m}= 0.3$,
$\Omega_{\Lambda}=0.7$ cosmology. The other reason is that the galaxy
luminosity functions that were used for older $dN/dz$ calculations had
a lower normalisation than the more recent estimates from large scale
optical and 21-cm surveys.

We emphasise that the normalisation of \fnhi\ and hence of $dN/dz$ is completely independent of the WHISP galaxy sample, but instead depends only on the \hi\ mass functions derived from HIPASS. This is a large scale blind 21-cm emission line survey 
covering the whole southern hemisphere and the redshift range $z=0$ to
$z=0.04$. Most of the weight to the normalisation of the \hi\ mass function 
comes from galaxies around $z=0.01$. Optical and infrared surveys have
shown that a large angular area around the southern Galactic pole is underdense 
by $\approx 25$ per cent extending out to $z=0.1$ \citep{frith2003,busswell2004}. 
This area occupies approximately one third of the HIPASS sky coverage, but it is not clear 
whether the underdensity extends to even larger regions in the southern hemisphere. 
In any case, our measured \fnhi\ and $dN/dz$ might be underestimated 
by up to 25 per cent due to this local galaxy deficiency. This implies that the
evolution in the comoving incidence rate is perhaps even slightly weaker than 
portrayed  in Fig.~\ref{dndz.fig}.

The lack of evolution in the comoving incidence rate since redshift
$z\sim1.5$ implies that the average \hi\ cross section above the DLA
limit has not changed significantly over half the age of the Universe.
At present, it is difficult to ascertain whether this should be
interpreted as no evolution in the DLA population -- meaning no change
in the space density {\em and\/} size of DLA absorbing systems -- or
as a combined effect where an evolving space density is compensated by
a change in mean absorber size.

%
\section{Expected properties of low-redshift DLA host galaxies}\label{properties.sec}
In this section we calculate the probability distribution functions of
the expected properties of galaxies responsible for high column
density \hi\ absorption at $z=0$.  To this end, we define a quantity
$\mathcal{N}$ as the `volume density of cross sectional area' for
different column density cut-offs. $\mathcal{N}$ is calculated as
\begin{equation}
\mathcal{N}({\bf y}) = \int \Sigma({\bf y},{\bf x}) \Phi({\bf x}) d{\bf x},
\end{equation}
where $\Sigma({\bf y},{\bf x})$ is the cross-sectional area of \hi\ in Mpc$^2$ above a certain column density cut-off as a function of galaxy properties ${\bf x}$ and  ${\bf y}$,
and $\Phi({\bf x})$ is the space density of galaxies in 
$\rm Mpc^{-3}$ as a function of ${\bf x}$. Similarly to the calculation of $f(\nhi)$, the parameter
${\bf x}$ has elements \mhi\ and $L$, such that $\Phi$ is again the \hi\ mass function or
the optical luminosity function that is used to calculate real space densities. The vector ${\bf y}$ could be any galaxy property, such as optical  surface brightness or morphological type.
For example, if ${\bf x}=\mhi$ and ${\bf y}=\mu$, we calculate $\mathcal{N}(\mu)$, the volume density of cross-sectional area as function of surface brightness $\mu$, using the \hi\ mass function to calculate space densities. In the remainder of this paper we will use ${\bf x}=\mhi$, so that 
the \hi\ mass function is used for calibration. We find that the conclusions would not change significantly if we were to use the optical luminosity function instead.

Another way of looking at $\mathcal{N}$ is that it defines the number
of systems above a certain column density limit that would be
encountered along a random 1 Mpc path through the $z=0$ Universe. Put
differently, $\mathcal{N}$ can be written as $dN/dz \times H_0/c$,
where $dN/dz$, the number of systems per unit redshift, is a familiar
quantity in QSO absorption line studies.  The run of $\mathcal{N}$ as
a function of galaxy property {\bf x} represents the probability
distribution of {\bf x} of galaxies responsible for absorption above a
certain column density.

In the following, we present this probability distribution for four
different column density limits, $\log \nhi>19.8$, $>20.3$, $>20.8$,
and $>21.3$.  The distribution for $\log \nhi=20.3$, which is the
classical DLA limit, is always shown as a thick line.  In table
\ref{relcon.tab} we tabulate the relative contributions of galaxies of
different luminosities and \hi\ masses to the total DLA cross
section. In table \ref{meanprop.tab} the mean and median properties of
expected DLA host galaxies are summarized.

\subsection{Luminosities}\label{lums.sec}
\begin{figure}\centering
\includegraphics[width=\columnwidth,trim=0.0cm 1.0cm 5.2cm 0.0cm]{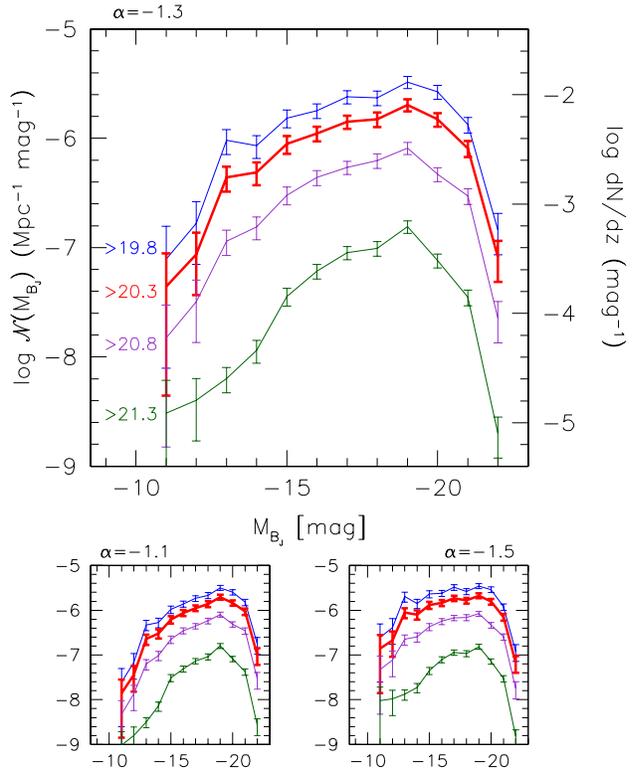}
    \caption{The expected distribution of absolute $B$-band magnitudes
    of high column density \hi\ absorbing systems. The lines plus
    errorbars show the product of cross sectional area and space
    density, which translates to the number of expected absorbers per
    Mpc per magnitude. The right axis shows the corresponding number
    of absorbers per unit redshift $dN/dz$.The different lines
    correspond to different column density limits, as indicated by the
    labels. The thick line corresponds to the classical DLA limit of
    $\log \nhi>20.3$.  The smaller panels show the effect of changing
    the slope of the low-mass slope of the \hi\ mass function that is
    used to calculate the normalisation. }
    \label{cross-mb.fig}
\vspace*{-0.25cm}\end{figure}

In Figure~\ref{cross-mb.fig} we show $\mathcal{N}$ as a function of
$B$-band absolute magnitude. As described in the preceding section, this 
distribution is calculated by multiplying the \hi\ mass function with the 
cross-sectional area  above a certain column density cut-off, and binning
in absolute magnitude.
The two lower panels in
Figure~\ref{cross-mb.fig} show the effect of changing the slope of the
HI mass function that is used to calibrate the distribution.  The luminosity
measurements are taken from the RC3, or when these are not available,
from LEDA, which gives $B$-band magnitudes transformed to the RC3
system. Since we use the $L_*$ measurement from \citet{norberg2002},
which is in the $b_J$ system, we convert our magnitudes to $b_J$ using
$b_J=B_{\rm RC3}+0.185$ as derived by \citet{liske2003}, and hence
Figure~\ref{cross-mb.fig} is approximately in the $b_J$ system.  What
is immediately obvious from this plot is that the probability
distribution is not strongly peaked around $L_*$ galaxies. Rather, for
column densities above the DLA limit, the distribution is almost flat
between $M_B\approx-15$ and $M_B\approx-20$.  The consequence of this
is that if an \hi\ column density $\nhi>10^{20.3} \,\icmsq$ is
encountered somewhere in the local Universe, the probability that this
gas is associated with an $L_*/50$ galaxy is only slightly lower than
for association with an $L_*$ galaxy.  More specifically, 87 per cent of the
DLA cross section is in sub-$L_*$ galaxies and 45 per cent of the cross
section is in galaxies with $L<L_*/10$.  These numbers agree very well
with the luminosity distribution of $z<1$ DLA host galaxies. Taking
into account the three non-detections of DLA host galaxies and
assuming that these are $\ll L_*$, we find that 80 per cent of the $z<1$ DLA
galaxies is sub-$L_*$.  The median absolute magnitude of a $z=0$ DLA
galaxy is expected to be $M_B=-18.1$ ($\sim L_*/7$), with 68 per cent in the
range $-15.5<M_B<-20.0$, whereas the mean luminosity is $L_*/2.5$.

By studying a sample of low-$z$ DLA galaxies, \citet{rao2003} also
conclude that low luminosity galaxies dominate the \hi\ cross section,
whereas \citet{chen2003} claim that luminous galaxies can explain most
of the DLA systems and that a contribution by dwarfs is not
necessary. The origin of this apparent disagreement lies in the
definition of a dwarf galaxy. Using the definition of \citet{rao2003}
that all sub-$L_*$ are dwarfs, these galaxies would indeed dominate
the DLA cross section. However, using the more stringent definition of
dwarf galaxies being fainter than $L_*/10$, these systems only
contribute approximately 45 per cent of the cross-section.

What can also be seen from Fig.~\ref{cross-mb.fig}, is that the
highest column densities ($\log \nhi>21.3$) are largely associated
with the most luminous galaxies: the probability distribution for the
highest column densities is more peaked around $L_*$. Small cross
sections of very high column density gas can still account for high
\hi\ masses, which explains why the total \hi\ mass density is
dominated by $L_*$ galaxies.

\subsection{\hi\ masses}
\begin{figure}\centering
\includegraphics[width=\columnwidth,trim=0.0cm 1.0cm 5.2cm 0.0cm]{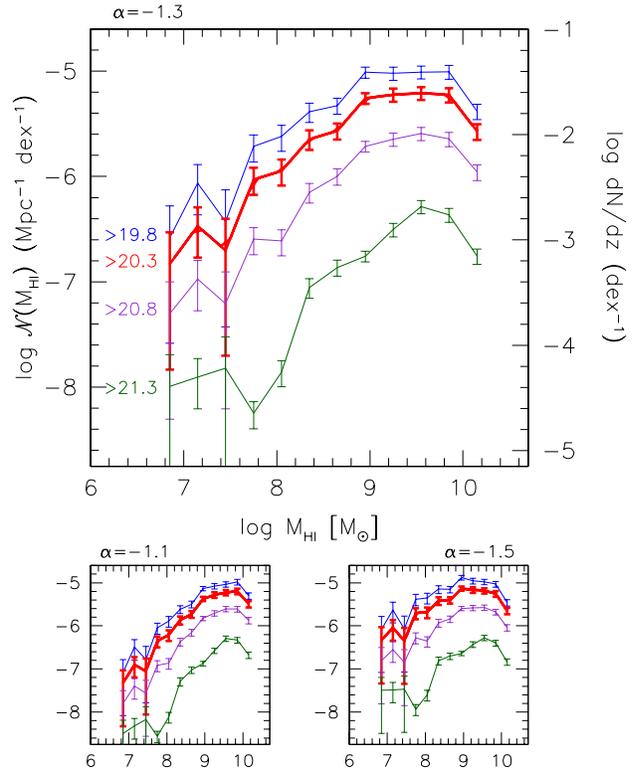}
    \caption{The expected distribution of the \hi\ masses of high
    column density \hi\ absorbing systems. The lines plus errorbars
    show the product of cross sectional area and space density, which
    translates to the number of expected absorbers per Mpc per decade
    of \hi\ mass. The right axis shows the corresponding number of
    absorbers per unit redshift $dN/dz$.The different lines correspond
    to different column density limits, as indicated by the
    labels. The thick line corresponds to the classical DLA limit of
    $\log \nhi>20.3$.  The smaller panels show the effect of changing
    the slope of the low-mass slope of the \hi\ mass function that is
    used to calculate the normalisation. }
    \label{cross-mhi.fig}
\vspace*{-0.25cm}\end{figure}

In Figure~\ref{cross-mhi.fig} we show $\mathcal{N}$ per decade of \hi\
mass as a function of log \hi\ mass. This figure shows largely the
same behaviour as Figure~\ref{cross-mb.fig}, but the distribution is
somewhat more peaked around \mhis\ galaxies
($\log\mhis/\msol=9.8$). The contribution from sub-$\mhis$ galaxies to
the total DLA cross section is still 81 per cent, that of galaxies with \hi\
masses lower than $10^9\msol$ is 31 per cent. These figures agree very well
with those of \citet{rosenberg2003} and \citet{ryan-weber2003}, based
on samples of galaxies observed with lower spatial resolution. The
median \hi\ mass of a $z=0$ DLA galaxy is expected to be $\log
\mhi/\msol=9.3$, with 68 per cent in the range $8.5<\log\mhi/\msol<9.8$.
Again we see that the highest column densities are preferentially
associated with the most massive galaxies. We will come back to this
fact in section~\ref{linking.sec}.

\subsection{Surface brightness and Hubble type}

\begin{figure}\centering
\includegraphics[width=\columnwidth,trim=0.0cm 6.5cm 5.2cm 0.0cm]{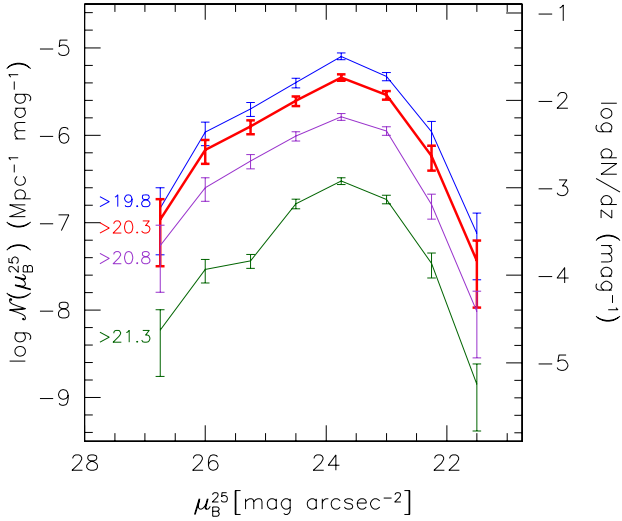}
    \caption{The expected distribution of the $B$-band surface
    brightness within the 25th mag isophote of host galaxies of high
    column density \hi\ absorbing systems. The lines plus errorbars
    show the product of cross sectional area and space density, which
    translates to the number of expected absorbers per Mpc per
    magnitude. The right axis shows the corresponding number of
    absorbers per unit redshift $dN/dz$.The different lines correspond
    to different column density limits, as indicated by the
    labels. The thick line corresponds to the classical DLA limit of
    $\log \nhi>20.3$.}
    \label{cross-mu.fig}
\vspace*{-0.25cm}\end{figure}

The probability distribution function of $\mu_B^{25}$, the mean
$B$-band surface brightness within the 25th mag arcsec$^{-2}$, is
given in Fig.~\ref{cross-mu.fig}. Similar to the finding of
\citet{ryan-weber2003}, we find that the cross section is dominated by
galaxies with $\mu_B^{25}$ in the range 23 to 24 mag
arcsec$^{-2}$. For reference, for our sample we find the median value
of $\mu_B^{25}$ for $L_*$ galaxies is 23.6 mag arcsec$^{-2}$. At the
bright end the distribution drops rapidly, showing that galaxies with
high surface brightnesses contain a small fraction of the cross
section. Toward dimmer galaxies, the distribution drops off slower. We
find that 53 per cent of the DLA cross section is in galaxies dimmer than
$\mu_B^{25} = 23.6$ mag arcsec$^{-2}$. However, for 8 per cent of the
galaxies in our sample no measurement of $\mu_B^{25}$ is
available. Assuming that these galaxies have no measurement because of
their low surface brightness, we are biased against LSB
galaxies. Including these in our lowest $\mu_B^{25}$ bins increases
the fraction of cross section in galaxies dimmer than $23.6$ mag
arcsec$^{-2}$ to 64 per cent.

The measurement of optical surface brightness that we have used tends
to understate the contribution of LSB galaxies to the DLA cross
section. The reason for this is that if the surface brightness is
measured within a fixed isophote, the radius of this isophote shrinks
if the surface brightness decreases. $\mu_B^{25}$ is therefore always
measured over the central brightest part of a galaxy, which naturally
decreases the dynamic range in surface brightness measurements. A
better alternative would be $\mu_B^{\rm eff}$, the effective surface
brightness (the average surface brightness within the half-light radius), 
but unfortunately this parameter is only available for
40 per cent of the galaxies in the WHISP sample. For illustrative purposes,
we can assign values of $\mu_B^{\rm eff}$ to those galaxies that have
no measurements, by using the surface brightness--luminosity relation
observed for nearby galaxies, for example by \citet{cross2002}. For
galaxies without $\mu_B^{\rm eff}$ we simply take the measurement of
the galaxy closest in absolute magnitude in our sample. Thus, we find
that the probability distribution of $\mu_B^{\rm eff}$ is much flatter
at the LSB end than that of $\mu_B^{25}$. Specifically, we find that
71 per cent of the cross section is in galaxies dimmer than $\mu_B^{\rm eff}
= 22.0$ mag arcsec$^{-2}$, which corresponds to the peak of the
distribution for $L_*$ galaxies. If an LSB galaxy is defined as having
a surface brightness fainter than 1.5 mag arcsec$^{-2}$ below this
value, we find that 44 per cent of the cross section is in LSB galaxies, in
accord with the findings of \citet{minchin2004}. The fractional
contribution of LSB galaxies to the DLA cross section is larger than
their contribution to the \hi\ mass density because their typical \hi\
mass densities are lower than those observed in high surface
brightness galaxies \citep[e.g.,][]{deblok1997}. This is supported by
the galaxy formation models of \citet{mo1998}, which indicate that
cross-section-selected samples are weighted towards galaxies with high
angular momentum (i.e., LSB galaxies).  In conclusion, our data is not
ideal to make firm statements about the contribution of LSB galaxies
to the DLA cross-section. However, using the information we have
available we can state that galaxies with surface brightness dimmer
than that of a typical $L_*$ galaxy make up at least half of the cross
section.

\begin{figure}\centering
\includegraphics[width=\columnwidth,trim=0.0cm 6.5cm 5.2cm 0.0cm]{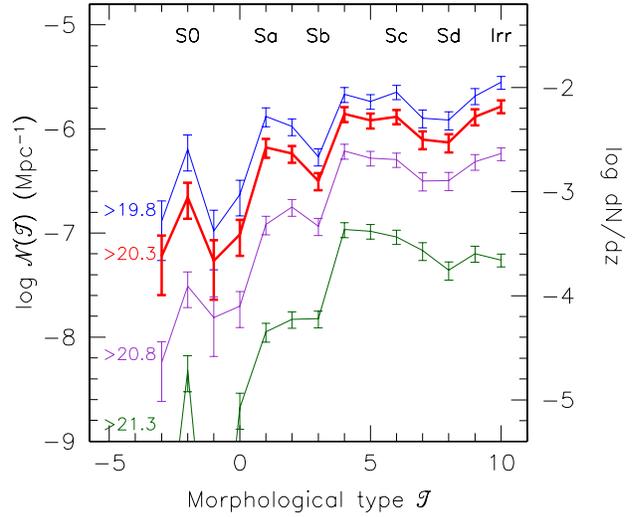}
    \caption{The expected distribution of the morphological types of
    host galaxies of high column density \hi\ absorbing systems. The
    lines plus errorbars show the product of cross sectional area and
    space density, which translates to the number of expected
    absorbers per Mpc. The right axis shows the corresponding number
    of absorbers per unit redshift $dN/dz$.The different lines
    correspond to different column density limits, as indicated by the
    labels. The thick line corresponds to the classical DLA limit of
    $\log \nhi>20.3$.  }  \label{cross-type.fig}
\vspace*{-0.25cm}\end{figure}

Turning now to the Hubble types, we see in Fig.~\ref{cross-type.fig}
that late-type galaxies are preponderant in the distribution of DLA
cross section. Earlier types are not negligible and the probability of
identifying an S0 galaxy with a DLA is approximately a third of
finding an Sc galaxy.

\begin{table}
\centering
\begin{minipage}{70mm}
\caption{Relative contribution to DLA cross section from different galaxies}
\label{relcon.tab}
\begin{tabular}{c c c c c}
\hline\hline
Quantity & $<L_*/10$ & $< L_*/5$ & $< L_*$ & $<2L_*$\\
\hline
$L$		& 0.45	& 0.58	& 0.87	& 0.96	\\
$\mhi$	& 0.22	& 0.37	& 0.81	& 0.96	\\
 \hline
\end{tabular}
\end{minipage}
\end{table}

\begin{table}
\centering
\begin{minipage}{70mm}
\caption{Expected properties of low-$z$ DLAs}
\label{meanprop.tab}
\begin{tabular}{l c c c}
\hline\hline
Quantity & median & mean & logarithmic mean\\
\hline\vspace{1mm}
$M_B$			& $-18.1_{-1.9}^{+2.6}$	&    -19.2	&    -17.7\\\vspace{1mm}
$\log \mhi$ ($M_\odot$)	& $9.3_{-0.7}^{+0.5}$	&	9.5	& 9.0\\\vspace{1mm}
$b$	(kpc)			&	$7.6_{-5.0}^{+10.2}$	&	10.6	&7.0\\
\hline
\end{tabular}
\end{minipage}
\end{table}

\subsection{Impact parameters}
\begin{figure}\centering
\includegraphics[width=\columnwidth,trim=0cm 0cm 0cm 0cm]{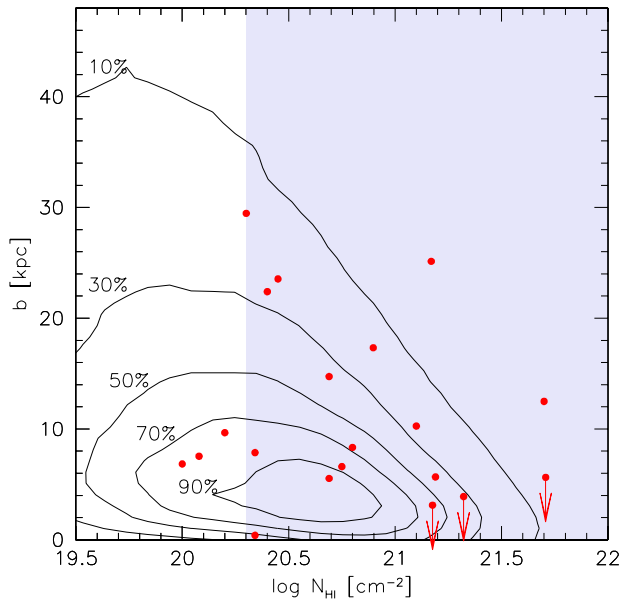}
    \caption{The two-dimensional probability distribution of \hi\
    cross section in the \hi\ column density - impact parameter
    plane. The contours are calculated from our \hi\ 21-cm maps and
    are drawn at 10, 30, 50, 70 and 90 per cent of maximum. The points are
    from DLA galaxy searches in the $z<1$ Universe. The shaded area
    shows the column density region corresponding to classical damped
    systems ($\log \nhi>20.3$)}
    \label{RNcontour.fig}
\vspace*{-0.25cm}\end{figure}

Figure~\ref{RNcontour.fig} shows the probability distribution of \hi\
cross section in the \nhi-$b$ plane, where $b$ is the impact parameter
in kpc from the position of the observed column density to the centre
of the galaxy. The WHISP sample was used to calculate the cross
sectional area contributed by each element $d\nhi db$ on a fine grid
in the \nhi-$b$ plane. We again used the type-specific \hi\ mass
functions to assign weights to each individual galaxy (see
section~\ref{fn.sec}). To increase the signal-to-noise in the figure,
we smoothed the probability distribution with a Gaussian filter, which
results in a final resolution of $\sigma=0.13$ dex in the \nhi\
direction and $\sigma=1.3$ kpc in the $b$ direction. The contour
levels are chosen at 10, 30, 50, 70, and 90 per cent of the maximum
value. The shaded area in the figure indicates the column density
region that corresponds to DLA column densities.

A few interesting features can be readily seen in
Figure~\ref{RNcontour.fig}.  The lowest contour illustrates that the
highest \hi\ column densities are very rarely seen at large
galactocentric radii: there appears to be a strong correlation between
\nhi\ and the maximum radius at which this is observed. The
observational fact that the \hi\ distribution in galaxies often shows
a central depression can be seen by the compression of contours near
$b=0$ kpc.  Finally, this plot shows that the largest concentration of
\hi\ cross section in galaxies in the local Universe is in column
densities in the range $20.3 < \log \nhi < 20.8$ and impact parameters
$b<7$ kpc.

The points in Figure~\ref{RNcontour.fig} are the pairs of \nhi\ and
$b$ measurements from the literature sample of low $z$ DLA galaxies
from Table~\ref{props.tab}. If low $z$ DLAs are drawn from the same
population of galaxies as those in our sample from the local Universe,
we would expect the points to show the same distribution as the
contours. Unfortunately, the statistics are too poor to calculate
contours from the DLA galaxy data, but there seems to a qualitative
agreement between the two data sets.

\begin{figure}\centering
\includegraphics[width=\columnwidth,trim=0cm 0cm 0cm 0cm]{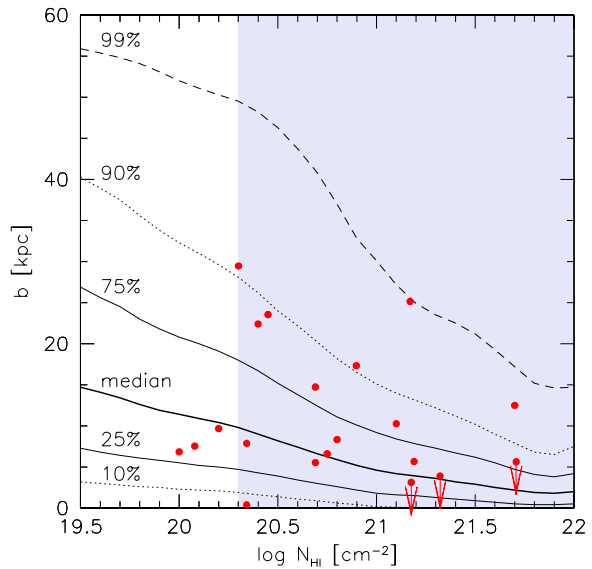}
    \caption{Conditional probability of impact parameter as a function
    of \hi\ column density.  The thick solid line shows the median
    impact parameter as calculated from our 21-cm emission line
    maps. The other lines show the 10th, 25th, 75th, 90th, and 99th 
    percentiles, as indicated by the labels. The points are from DLA galaxy
    searches in the $z<1$ Universe. }
    \label{condprobRN.fig}
\vspace*{-0.25cm}\end{figure}

A more straightforward comparison between the DLA galaxies and local
galaxy properties is presented in Figure~\ref{condprobRN.fig}, which
shows the conditional probability distribution of impact parameter $b$
as a function of column density \nhi. This figure is calculated by
normalising at each value of \nhi\ in Figure~\ref{RNcontour.fig} the
distribution function of cross section as function of $b$ to the peak
of the distribution.  The solid line shows the median $b$ value as a
function of \nhi, the other lines show the 10th, 25th, 75th, 90th, and 99th
percentiles, respectively. The literature values are again overplotted as
points.  Both in the $z=0$ data and in the DLA data, there is a weak
correlation between \nhi\ and $b$.  \citet{rao2003} also noted the
existence of this relation. The agreement between the points and the
contours is remarkably good: 50 per cent of the literature values are within
the 25th and 75th percentiles, and 75 per cent are within the 10th and 90th percentile contours.  Based on
our analysis, the expected median impact parameter of $\log \nhi>20.3$
systems is 7.6 kpc, whereas the median impact parameter of identified
$z<1$ DLA galaxies is 8.3 kpc. 
Although we are limited by small number statistics, a comparison 
between the contours and the points suggests that the observed number
of very low $b$ systems is lower than expected. 
These identifications are the ones most likely to be missed due to the
proximity of a the bright background QSO. Alternatively, if dust obscuration 
in DLA host galaxies is important, these low $b$ sight lines 
are expected to be the first to drop out of a flux limited quasar sample.

\begin{figure}\centering
\includegraphics[width=\columnwidth,trim=0.0cm 6.5cm 5.2cm 0.0cm]{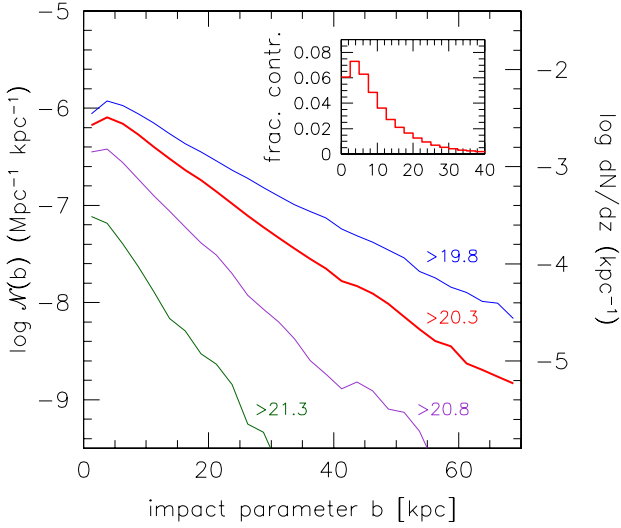}
        \caption{The probability distribution of impact parameter $b$
        between the background QSO and the centre of a galaxy giving
        rise to \hi\ absorption. The lines show the product of cross
        sectional area and space density, which translates to the
        number of expected absorbers per Mpc per kpc. The right axis
        shows the corresponding number of absorbers per unit redshift
        $dN/dz$.The different lines correspond to different column
        density limits, as indicated by the labels. The thick line
        corresponds to the classical DLA limit of $\log \nhi>20.3$.
        The inset shows the probability distribution of $b$ for column
        densities $\log \nhi>20.3$ on a linear vertical scale.  }
        
    \label{cross-radius.fig}
\vspace*{-0.25cm}\end{figure}

A more detailed view of the distribution of impact parameters is
presented in Figure~\ref{cross-radius.fig}, which shows the predicted
probability distribution of $b$ for different column density
cut-offs. The peak of the distribution for DLA column densities is at
$b\sim 5$ kpc and the distribution drops rapidly toward higher values
of $b$. The inset shows the probability distribution of $b$ above the
DLA limit on a linear scale.  For higher column densities, the
probability distributions drop off even more rapidly, which again
shows that it is extremely unlikely to encounter a high \nhi\ at a
large separation from the centre of a galaxy. For DLA column
densities, we find that 60 per cent of the host galaxies are expected at
impact parameters $<10 \,\rm kpc$ and 32 per cent at $<5 \,\rm kpc$. Assuming
no evolution in the properties of galaxy's gas disk, these numbers
imply that 37 per cent of the impact parameters are expected to be less than
$1''$ for systems at $z=0.5$ and 48 per cent less than $1''$ at $z=1$. These
numbers illustrate that very high spatial resolution imaging programs
are required to successfully identify a typical DLA galaxy at $z\sim
1$.

The intrinsic assumption in this comparison is that the low $z$ DLA
galaxy sample is a fair cross-section selected sample. In reality the
sample is a compilation of many surveys, using different selection
techniques, and different resolutions and wave bands to image the
galaxy. Keeping this limitation in mind, we conclude that the measured
impact parameters and column densities of low $z$ DLA galaxies is in
agreement with the hypothesis that DLA galaxies can be explained by
the local galaxy population.

\begin{figure}\centering
\includegraphics[width=\columnwidth,trim=0.cm 10.0cm 0cm 0cm]{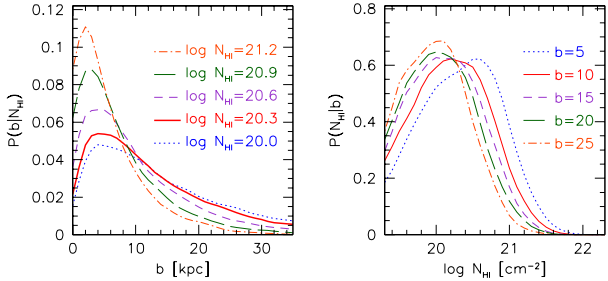}
    \caption{{\em Left\/:} Normalised probability distribution of
    impact parameter $b$ for different column densities. For \hi\
    column densities near the DLA limit of $2\times 10^{20} \, \rm
    cm^{-2}$, the most likely impact parameter to the host galaxy is
    $5\, h_{75}^{-1}$ kpc. {\em Right\/:} Normalised probability
    distribution of \hi\ column density for different impact
    parameters $b$.}
    \label{prop.fig}
\vspace*{-0.25cm}\end{figure}

For completeness, we show in Figure~\ref{prop.fig} the expected
probability distribution function of impact parameter $b$ {\em at\/}
various column densities. Note the difference with the previous
diagrams, where we plotted probability distribution functions {\em
above\/} certain column density cut-offs. The right panel is the
probability distribution function of column density at various values
of $b$.

\subsection{Linking DLA parameters to galaxy properties}\label{linking.sec}
\begin{figure}\centering
\includegraphics[width=\columnwidth,trim=0.cm 0.0cm 0cm 0cm]{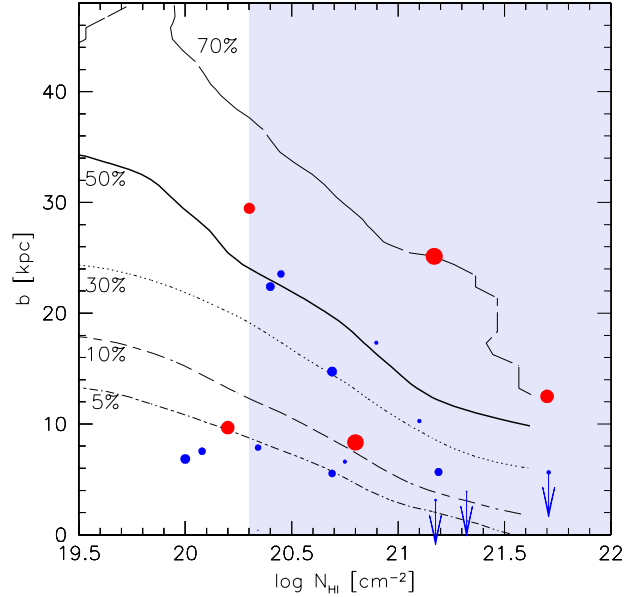}
    \caption{Probability distribution of optical luminosities of
    galaxies giving rise to \hi\ absorption with column density \nhi\
    and impact parameter $b$. The lines indicate the probability that
    the host galaxy is more luminous than an $L_*$ galaxy, and represent
    5, 10, 30, 50 (thick line) and 70 per cent probabilities, from bottom to
    top. The points are from DLA galaxy searches and are the same as
    in Fig.~\ref{RNcontour.fig} and Fig.~\ref{condprobRN.fig}, but
    here the symbol size represent intrinsic luminosity of the host
    galaxy. Red points indicate galaxies brighter than $L_*$, blue
    points are fainter than $L_*$.}
    \label{L_prob.fig}
\vspace*{-0.25cm}\end{figure}

We conclude our comparisons between DLA systems and local galaxies by
looking at the combined probability distribution function of column
density, impact parameter and host galaxy luminosity.  In
Fig.~\ref{L_prob.fig} we again show the \nhi\ and $b$ measurements
from low-$z$ DLA galaxies from the literature, but this time the
symbol size reflects the luminosity of the galaxies such that the
symbol area scales in direct proportion to $L/L_*$. The contours
represent the probabilities that the combined measurement of $\nhi$
and $b$ is expected for a galaxy with luminosity $L>L_*$. For example,
the thick solid line divides the diagram in two regions, above this
line most host galaxies would be more luminous than $L_*$, below this
line most would be less luminous than $L_*$. Apparently, the most
luminous galaxies are most likely associated with high column density
DLAs, at large impact parameters from the background QSOs.

\begin{figure}\centering
\includegraphics[width=\columnwidth,trim=0.cm 0.0cm 0cm 0cm]{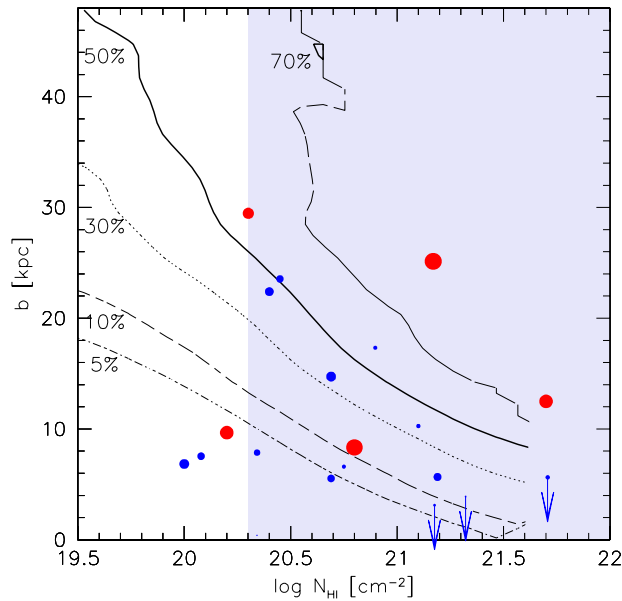}
    \caption{Probability distribution of \hi\ masses of galaxies
    giving rise to \hi\ absorption with column density \nhi\ and
    impact parameter $b$. The lines indicate the probability that the
    host galaxy has an \hi\ mass in excess of $\mhi^*$, and represent
    5, 10, 30, 50 (thick line) and 70 per cent probabilities, from bottom to
    top. The points are the same as in Fig.~\ref{L_prob.fig}}
    \label{MHI_prob.fig}
\vspace*{-0.25cm}\end{figure}

The probability contours can be directly compared to the distribution
of symbol sizes. Although the scatter is large, we see a general
agreement between the points and the contours: below the 5 per cent line
mostly low luminosity galaxies are found, and on the other hand, the
two galaxies above the 70 per cent line are both galaxies with high
luminosities. A similar diagram is presented in
Fig.~\ref{MHI_prob.fig}, but here contours indicate the probability
that the host galaxy has an \hi\ mass in excess of \mhis. The
conclusions from this figure are very similar as those from
Figure~\ref{L_prob.fig}.

\subsection{Comparison to models and simulations of DLAs}
\begin{figure}\centering
\includegraphics[width=\columnwidth,trim=0.cm 0.0cm 0cm 0cm]{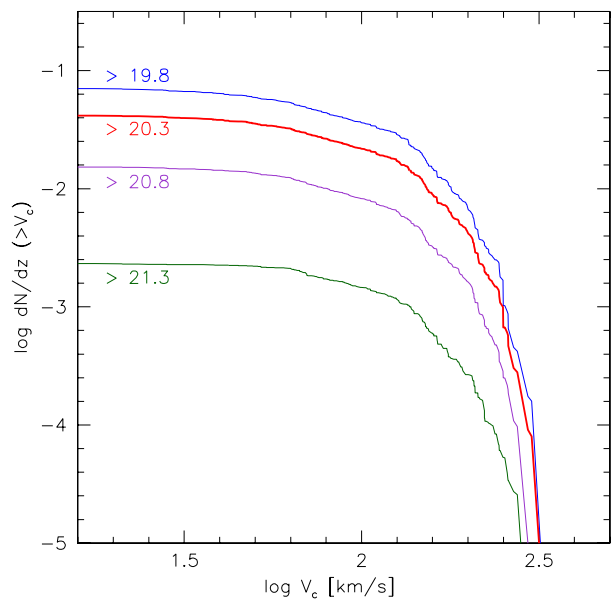}
    \caption{The cumulative redshift number density of DLAs in
    galaxies of different rotational velocities. The different lines
    correspond to different column density limits, the thick line is
    for $\log\nhi>20.3$. }
    \label{dndz_vc.fig}
\vspace*{-0.25cm}\end{figure}

In order to compare our findings with numbers calculated in models of
galaxy formation, we first transform the $B$-band luminosities of our
WHISP galaxies into rotational velocities via the Tully-Fisher
relation.\footnote{In principle, a measurement of rotational velocity
is available directly from the WHISP data, but at present this
analysis has not been completed for the full sample. Since the
observed scatter in the Tully-Fisher relation is very small, the
conclusions will not change by using this approximation.} We choose
here to use the Tully-Fisher relation determined by \citet{meyer2005},
based on the HIPASS sample.  Figure~\ref{dndz_vc.fig} shows the
cumulative distribution of DLA redshift number density for galaxies
with different rotational velocities $V_{\rm circ}$.  We present the
cumulative distribution because this representation is normally used
in publications based on cosmological simulations, and can therefore
be directly compared to those.

On the basis of semi-analytical models \citet{okoshi2005} find that
the average virial velocity of a $z=0$ DLA is $V_{\rm vir}\sim 90\,
\kms$ and the average luminosity is $0.05L_*$. In their models, the
fraction of DLA hosts in galaxies fainter than $L_*/10$ is 98 per cent,
whereas we find that this fraction is only 41 per cent. Furthermore, they
find that the typical impact parameter is 3 kpc, much smaller that our
median value of 7.8 kpc. To make their results agree with observations
of low-$z$ DLA galaxies, \citet{okoshi2005} propose that the masking
effect where the DLA galaxies are contaminated by the point spread
function of the QSO, hinders the identification of 60 per cent to 90 per cent of DLA
galaxies with small impact parameters.  The SPH simulations of
\citet{nagamine2004} also suggest that the relative contribution of
low mass galaxies to the DLA cross section is very large: their
cumulative contribution of redshift number density as a function of
$V_{\rm circ}$ is steeper than what we find in Figure
\ref{dndz_vc.fig}. These authors note that the $z=0$ results should be
taken with caution because the mass resolutions that these results are
based on are low and higher resolution SPH simulations are probably
required to arrive at more precise results at $z=0$.  Other work, such
as that of \citet{gardner2001}, \citet{mo1998,mo1999},
\citet{haehnelt2000}, and \citet{maller2001} mostly concentrate on the
high redshift ($z\sim 3$) DLA population and cannot be compared
directly to our work, but also point to sub-$L_*$ galaxies as the major
contributors to the DLA cross section.  In conclusion, semi-analytical
models and cosmological simulations generally over-predict the redshift
number density contribution of low mass systems. A notable exception
is the simple models of \cite{boissier2003}, which show that the peak
and the median of the cross section distribution lies around $L_*$
galaxies. These models under-predict the importance of low mass
systems, both in comparison with our results and in comparison with
low $z$ DLA host galaxies.

From Figure~\ref{dndz_vc.fig} we find that the mean rotational
velocity $V_{\rm circ}$ of a $z=0$ DLA is $111\, \kms$ (the
log-weighted mean $V_{\rm circ}$ is $95\,\kms$).  Note that the
rotational velocities we measure are the peak velocities of the
galaxies' rotation curves, whereas in simulations galaxies are
normally characterised by their virial velocity $V_{\rm vir}$. The
ratio between these two values depends on the concentration index of
the dark halo, but a typical value is $V_{\rm circ} \approx 1.4V_{\rm
vir}$ \citep[see][]{bullock2001}, which implies that the log-weighted
mean virial velocity of a $z=0$ DLA is approximately $70
\,\kms$. Using the relation between $V_{\rm circ}$ and the virial mass
$M_{\rm vir}$ given by \citet{bullock2001}, we find that the mean
total mass of a $z=0$ DLA would be $\approx 1.5\times 10^{11}\msol$.

%
\section{Metal abundances of low redshift DLAs} \label{metal.sec}
In the previous section we have presented evidence that DLAs arise in the gas disks of galaxies such as those in the $z=0$ population. Measuring metallicities in DLAs therefore should provide information on abundances of the interstellar matter in galaxies. Determining DLA metallicities as a function of redshift will probe the history of metal production in galaxies over cosmic time. However, the metallicities typically measured in DLAs are low (around 1/13 solar), which is often taken as an indication that DLAs do not trace the general galaxy population, for which a mass-weighted mean metallicity of near-solar is expected at low redshift \citep[see e.g.,][]{pettini1997}. In this section we present a more detailed analysis of the expected metallicities of DLAs, under the assumption that they arise in the gas disks of normal present day galaxies. We have no direct measurements of metal abundances for the complete sample of WHISP galaxies. However, we can make use of metallicity studies in other galaxies to statistically assign metallicities to our sample. Although this approach might introduce some degree of uncertainty in the results, it will help in understanding the observed metal abundance measurements in DLA systems. It should be kept in mind that the abundances of local galaxies are measured from emission lines arising in the photo-ionised gas, while the measurements in DLAs are from absorption lines in the neutral gas. However, recent work by \citet{schulte-ladbeck2005} shows that at least in one well-studied nearby galaxy the emission and absorption measurements of the same $\alpha$-elements are consistent.
\subsection{The expected metallicities of DLAs at $z=0$}
We take two different approaches to assigning metallicities to our
sample of galaxies.  The first approach is to adopt the
well-established metallicity-luminosity ($Z-L$) relation for local
galaxies and apply that to our sample. We choose to adopt the relation
given by \citet{garnett2002} for oxygen abundances:
\begin{equation}
\log{\rm (O/H)}= - 0.16 M_B - 6.4.
\label{oh-lum.eq}\end{equation}
The results do not change significantly if instead we use the more
recent $Z-L$ relation derived from SDSS imaging and spectroscopy of
53,000 galaxies \citep{tremonti2004}.  
To convert the log(O/H) values to solar values we adopt the solar
abundance of $12+\log{\rm (O/H)}=8.66$ from \citet{asplund2004}.  The
effect of abundance gradients as a function of galactocentric distance
is taken into account by using the result from \cite{ferguson1998},
who found a mean gradient of [O/H] of $-0.09 \rm \,dex\, kpc^{-1}$
along the major axes of spiral galaxies. We assume that this gradient
is equal for all galaxies and that it does not vary as a function of
radius. We apply a simple geometrical correction to correct for
inclined disks. Furthermore, we assume that Eq.~\ref{oh-lum.eq}
applies to the mean abundance measurement within $R_{25}$, which
agrees with \citet{garnett2002}. For each galaxy in our sample we
scale the offset of the gradient such that the mean abundance
measurement within $R_{25}$ fits the metallicity-luminosity relation.
This approach intrinsically assumes that the [O/H] abundance gradients
are independent of galaxy luminosity. To test the effect of this
assumption on the calculation of expected metallicities of DLAs, we
also take a second approach in which we adopt an [O/H] gradient that
has both the intercept and slope varying with galaxy absolute
magnitude. We fit straight lines to the relations found by
\citet{vila-costas1992} (their Figure 11) to find the varying slope
and intercept. The two approaches depend on completely independent
observations of metal abundances in local galaxies.

\begin{figure}\centering
\includegraphics[width=\columnwidth,trim=0.0cm 6.5cm 5.2cm 0.0cm]{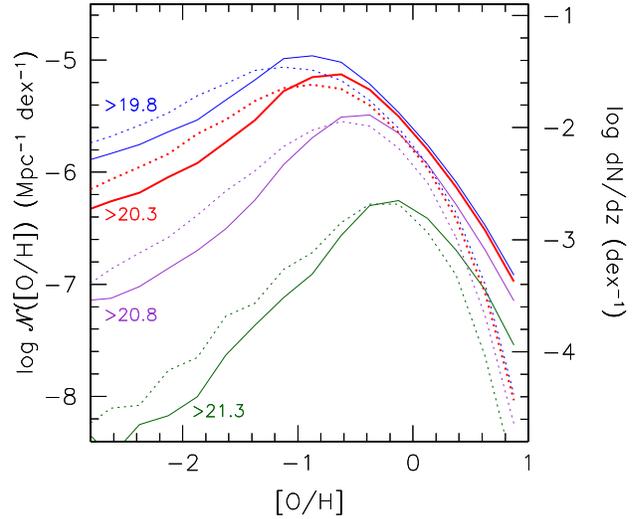}
        \caption{The probability distribution of oxygen abundance
[O/H] of \hi\ absorbers. The lines show the product of cross sectional
area and space density, which translates to the number of expected
absorbers per Mpc per decade. The right axis shows the corresponding
number of absorbers per unit redshift $dN/dz$.  The solid lines refer
to the approach of assuming fixed [O/H] gradients and the the dashed
lines refer to varying gradients (see text).  The four different lines
for each approach correspond to different column density limits, as
indicated by the labels. The thick lines corresponds to the classical
DLA limit of $\log \nhi>20.3$.  }
    \label{cross-abun.fig}
\vspace*{-0.25cm}\end{figure}

Every pixel in our 21-cm maps is assigned [O/H] values, based on the
two procedures.  Applying a similar method to that set out in
section~\ref{properties.sec}, we can now calculate the expected [O/H]
distribution for cross-section selected samples, for different column
density limits. The results are shown in Figure~\ref{cross-abun.fig},
where the solid lines refer to the first approach of fixed gradients,
and the dashed lines refer to the varying gradients.

Although the distributions peak at slightly different locations, it is
obvious that the global trends are not strongly dependent on our
assumptions on abundance gradients in disks.  The main conclusion is
that the metallicity distribution for \hi\ column densities $\log
\nhi>20.3$ peaks around [O/H]=$-1$ to $-0.7$, much lower than the mean
value of an $L_*$ galaxy of [O/H]$\approx 0$.  Furthermore, we see a
strong correlation between the peak of the [O/H] probability
distribution and the \hi\ column density limit: for \hi\ column
densities in excess of $\log \nhi=21.3$, the expected peak of the
distribution is at [O/H]=$-0.2$. This correlation is expected, since
our assumed radial abundance gradients impose a relation between \nhi\
and [O/H].  Taking into account uncertainties in the \hi\ mass
function, in the $L-Z$ relation and in the abundance gradients in
galaxies, we adopt a value of [O/H]$=-0.85\pm 0.2$ as a representative
value for the median cross-section--weighted abundance of \hi\ gas in
the local Universe above the DLA column density limit.

\subsection{Comparison to low-$z$ DLA metallicity measurements}
It is interesting to compare this expected abundance of DLAs at $z=0$ to higher redshift values to see whether the measurements can be brought into agreement. The first evidence of evolution in DLA metallicities over cosmic time was presented by \citet{kulkarni2002}, contrary to previous claims of no evolution \citep[e.g.,][]{pettini1997,pettini1999}. This results was further established by an analysis of 125 DLA metallicity measurements over the redshift range $0.5<z<5$ by \citet{prochaska2003}, who reported significant evolution of $-0.26\pm 0.07\,\rm dex$ per unit redshift in the mean metallicity [M/H]. \citet{kulkarni2005} compiled new metallicity measurements in four DLAs at $z<0.52$ and combined with literature measurements this study also finds evolution in [M/H] of $-0.18 \pm 0.07\,\rm dex$ per unit redshift. The \citet{prochaska2003} metallicity measurements are mostly based on $\alpha$-element abundances, whereas the \citet{kulkarni2005} data is mostly based on Zn abundances. We refer to \citet{kulkarni2005} for a discussion on why Zn is an appropriate choice as a metallicity indicator.

Before we compare our $z=0$ results to those for DLAs at higher
redshifts, we discuss the different measurements of `mean metallicity'
that we can use to make the comparison.  The first is to calculate the
global interstellar metallicity $\bar{Z}$, defined as 
\begin{equation}
\bar{Z}=\frac{\int \nhi \fnhi Z(\nhi) d\nhi}{\int \nhi \fnhi d\nhi}.
\end{equation}
The reason that this equation gives a true measurement of the 
interstellar metallicity is that absorption lines are cross-section--selected, which implies that a correct weighting of different regions in absorbing systems is automatically
taken into account, as is stressed by \citet{kulkarni2002}.
In practice,  in DLA studies $\bar{Z}$ is
normally calculated by taking the average over the metallicities of
individual absorption-line systems weighted by their \hi\ column
densities.
For our
low $z$ data we can determine $\bar{Z}$ by simply taking the
mass-weighted average of all mean metallicities of the WHISP galaxies:
\begin{equation}
\bar{Z}=\frac{\int \Theta(\mhi) \mhi Z(\mhi) d\mhi}{\int \Theta(\mhi) \mhi d\mhi},
\end{equation}
where $\Theta(\mhi)$ is the \hi\ mass function and $Z(\mhi)$ is the
mean metallicity of a galaxy with \hi\ mass \mhi. We can calculate
$Z(\mhi)$ by taking the column density-weighted metallicity of each
WHISP galaxy, but only counting column densities above the DLA
limit. Applying this method we find values of $\bar{Z}=0.58 Z_\odot$ and
$\bar{Z}=0.35 Z_\odot$ for the fixed gradient and the variable gradient
method, respectively.  \citet{kulkarni2002} and \citet{fukugita2004}
applied similar techniques, but integrated over the optical luminosity
function and found values of $\bar{Z}\approx 0.8 Z_\odot$ and $\bar{Z}=0.83\pm
0.25 Z_\odot$.  The reason for these values being higher than ours is that
these studies use one global metallicity measurement for each galaxy (which 
involves some radial averaging over the inner parts of galaxies), while our analysis
takes into account radial abundance gradients over the whole \hi\ gas disk. 
For the mean mass-weighted metallicity of \hi\ gas
with $\log \nhi>20.3$ at $z=0$ we adopt the value of
$\log \bar{Z}/Z_\odot=-0.35\pm 0.2$, where the errors again represent all
uncertainties in arriving at this result.

The second measurement of metallicity is the cross-section--weighted
average, which we will refer to as $\hat{Z}$. For DLAs, this
measurement is simply the unweighted mean of the metallicities and
represents a `typical' metallicity of DLAs.  For our $z=0$ sample, we
can make use of the calculations presented in
Fig.~\ref{cross-abun.fig}, which shows the cross-section--weighted
[O/H] distribution of high column density \hi\ gas in the local
Universe.  We find values of $\hat{Z}=0.36 Z_\odot$ and $\hat{Z}=0.21 Z_\odot$ for the
fixed gradient and the variable gradient method, respectively, and
hence adopt a value of $\log \hat{Z}/Z_\odot=-0.55\pm 0.2$, for the `typical'
metallicity of $z=0$ DLAs.

The third method stems directly from the way we presented the
probability distribution of metallicity in
Figure~\ref{cross-abun.fig}. As discussed above, we derive from this
the cross-section--weighted median metallicity, for which we adopted
$\log\tilde{Z}/Z_\odot=-0.85\pm 0.2$.

\begin{figure}\centering
\includegraphics[width=\columnwidth,trim=0cm 0cm 0cm 0cm]{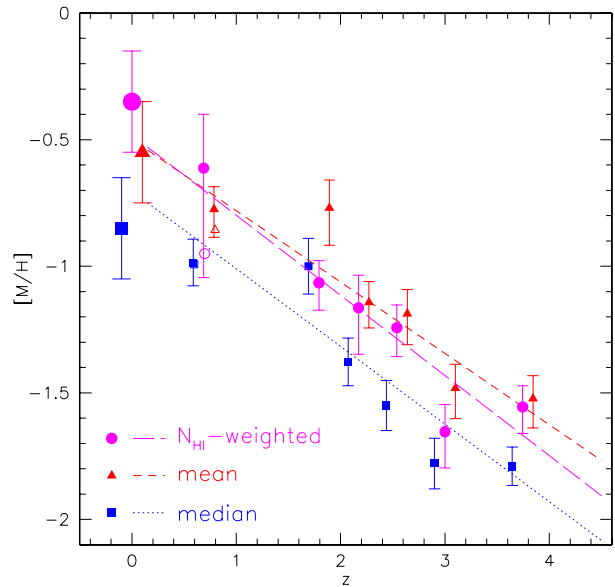}
    \caption{The metallicity [M/H] measurements in DLAs as a function
     of redshift. Data at redshifts $z>0$ are taken from \citet{prochaska2003} and \citet{kulkarni2005}. Survival analysis has been used to calculate mean and
     median values because the data include many upper limits. 
     The symbols refer to column density-weighted mean (filled circles),
     the cross-section--weighted mean (filled triangles), and median
     values (filled squares) in six redshift bins.  The errorbars indicate
     1 $\sigma$ statistical uncertainties. The lines are least squares fits to the 
      $z>0$ data points. The points at $z=0$ are from our analysis as described in 
      the text. The open symbols at $z\sim 0.6$ show the effect of excluding the 
     X-ray metallicity measurement in Q0235+164.  
      The squares and triangles are offset horizontally by $-0.1$ dex
     and $0.1$ dex, respectively.  }
    \label{Zz.fig}
\vspace*{-0.25cm}\end{figure}

In Figure~\ref{Zz.fig} we show the metallicities in DLAs as a function of redshift, extracted from the combined compilations by \citet{prochaska2003} and \citet{kulkarni2005}. In amalgamating the two data sets we gave preference to Zn measurements for systems that occurred in both lists. The \citet{kulkarni2005} data include many Zn measurements at low $z$, but a large fraction of those are upper limits. In order to take these limits into account appropriately, we applied the Kaplan-Meier estimator for randomly censored data sets, as implemented in the survival analysis package ASURV \citep{feigelson1985}. In six different redshift bins (approximately the same as those defined by \citealt{kulkarni2005}) we calculate the mean \nhi-weighted ($\bar{Z}$), the mean cross-section--weighted ($\hat{Z}$) and the median metallicities, as indicated by solid circles, triangles and stars, respectively. The errorbars are 1$\sigma$ statistical uncertainties as given by the Kaplan-Meier estimator. The long-dashed, short-dashed and dotted lines show the linear least squares fits to each set of measurements. The larger symbols at $z=0$ show our estimates of metallicities in the local Universe.

For all measurements of metallicity a slope between $-0.25$ and $-0.3$ dex per unit redshift can be seen, in agreement with the findings of \citet{prochaska2003} and \citet{kulkarni2005}, on whose compilations of data this figure is based. This increase in metallicity is consistent between the highest redshift points at $z\sim 4$ and the lowest points at $z\sim 0.6$. If we assume that DLAs arise in the gas disks of galaxies, where continued star formation causes a release of metals into the ISM, it seems reasonable to assume that this evolution in metallicity persists down to $z=0$. Note also that most models of cosmic chemical evolution predict a nearly exponential increase in metallicity as a function of redshift \citep[c.f.,][]{kulkarni2002}. Given this assumption, we find from Figure~\ref{Zz.fig} that the expected metallicity of DLAs at $z=0$ is in excellent agreement with our estimates from the local galaxy population: the $z=0$ extrapolations of all three lines are within the uncertainties of our estimates.

 \citet{kulkarni2005} have argued that the \nhi-weighted DLA metallicities do not rise up to near-solar values at $z=0$. These authors find a DLA metal
abundance of $-0.79\pm 0.18 Z_\odot$ at $z=0$ using survival analysis, a
factor of two lower than the \citet{prochaska2003} extrapolation. This
discrepancy lies mostly in the exclusion of the x-ray absorption based
metallicity of the low-$z$ system AO 0235+164 \citep{junkkarinen2004}. This system causes the large errorbars on the $z\sim 0.6$ points in
Figure~\ref{Zz.fig}. The open symbols in Fig.~\ref{Zz.fig} shows the effect of
excluding AO 0235+164. The mean and median [M/H] are not affected much, but the
\nhi-weighted value drops by more than a factor of two because of the high 
\nhi\ measured in this system. Hence, 
if we only used the UV absorption measurements, we
would find that our \nhi-weighted $\bar{Z}$ in galaxies is a factor
two higher than the extrapolated value from DLAs.
Different authors disagree on whether it is fair
to include this system \citep[c.f.,][]{chen2005,kulkarni2005}, but at least
it illustrates how poor the statistics are on low-$z$ abundance
measurements. The lowest redshift bin includes 17 measurements (including AO 0235+164), of which  five are limits. For illustration, if we calculate 
$\log\bar{Z}/Z_{\odot}$ at $z=0$ from 17 random sightlines through our galaxy sample, we find that the statistical error on this value is $-0.2+0.35$ dex, demonstrating that metallicity measurements based on small samples of DLAs are unavoidably marked by large uncertainties \citep[see also][]{chen2005}. 

Interestingly, we see that for local galaxies the \nhi-weighted
$\bar{Z}$ is higher than the mean $\hat{Z}$. Although the statistics are poor, 
it appears that for $z>1.5$ DLAs this
is not the case. The origin of this difference might lie in the dust
obscuration bias in DLAs: DLAs with the highest \hi\ column densities
and highest metallicities could be missed in magnitude-limited surveys
as a consequence of their own extinction \citep[see e.g.][for a
continued discussion on this issue]{fall1989,
fall1993,ellison2001,murphy2004,vladilo2005,wild2005}. If, like at $z=0$, there
is an intrinsic positive correlation between \nhi\ and metallicity,
this bias would have a much higher effect on the \nhi-weighted
$\bar{Z}$ than on the mean $\hat{Z}$ of DLAs, causing the two
measurements to be very similar in magnitude. In our local galaxy
sample such a bias would not exist. The fact that we see a factor of
two difference between the two values at $z=0$ and not for $z>1.5$ DLAs
therefore is suggestive of dust obscuration introducing biases in DLA
samples.

Finally, we comment on the scatter in the observed metallicity
measurements in DLAs.  From the probability distribution of [O/H] in
Fig.~\ref{cross-abun.fig} we find that the expected 1$\sigma$
scatter in cross section-selected metallicity measurements is 0.7
dex. A similar value is estimated by \citet{chen2005}.  This is
slightly larger than the typical scatter observed in metallicity
measurements at high redshift of $\sim 0.5$ dex, illustrating that
metallicity gradients in galaxy disks can easily explain the observed
scatter in DLA metallicity measurements.

In conclusion, we find that the cross-section--weighted mean
metallicity of local galaxies as well as the mass-weighted metallicity
are in very good agreement with the hypothesis that DLAs arise in the
\hi\ disks of galaxies, provided that the metallicities in DLAs
continue to evolve since $z=0.5$ with a rate similar to that observed
between $z=0.5$ and $z=4$.

Before ending this section, we wish to comment briefly on the work of
\citet{chen2005}, who presented similar conclusions based on
Monte-Carlo simulations.  These authors use average radial \hi\
profiles for three types of galaxies and assume that the metallicity
gradients in these galaxies are similar to what is found for their
sample of DLAs. Their analysis does not take into account that
galaxies over a large range in absolute magnitude contribute to the
DLA cross section (see section \ref{lums.sec}), and effectively, only
$L_*$ galaxies are considered. Therefore, the $L-Z$ relation is not
taken into account in this analysis.  Furthermore, in their analysis,
the mean metallicity is measured over the whole gas disk out to the
radius where 21-cm emission is normally detected, and not just over
the region $\log \nhi>20.3$.  Since there is an intrinsic relation
between \nhi\ and metallicity, this causes an underestimation of the
mean $Z$ for DLAs. This explains why the \citet{chen2005} estimate of
the \nhi-weighted mean $\bar{Z}=0.3 Z_\odot$ 
is lower than our estimate.


\section{Conclusions}
In this paper we tested the hypothesis that DLA absorption lines
observed in the spectra of background QSOs arise in gas disks of
galaxies like those in the $z=0$ population.  Since the \hi\ column
densities seen in DLA systems ($\nhi>2 \times 10^{20}$) are the same
as those routinely observed in 21-cm emission line studies of local
galaxies, we can make use of these observations to test the
hypothesis. Thus, we used a sample of 355 high quality WSRT 21-cm
emission line maps to calculate in detail the expected column density
distribution function, the redshift number density, and the expected
probability distribution functions of different galaxy parameters of
the low redshift DLAs.  We summarise the conclusions as follows:

1) The local galaxy population can explain the incidence rate of low
redshift DLAs. There appears to be no evolution in the `cross section
times space density' or the mean free path between absorbers from
$z\sim 1.5$ to $z=0$. Between the highest redshifts at which DLAs are
found ($z\sim 4-5$) and the present time, the evolution in comoving
incidence rate is only approximately a factor of two.  We find that
$dN/dz(z=0)=0.045 \pm 0.006$.

2) Based on the local galaxy population it is expected that the DLA
cross section is dominated by sub-$L_*$ galaxies ($87$ per cent). This agrees
with the statistics of identified DLA host galaxies at low
redshifts. 50 per cent of the low redshift DLAs should arise in galaxies with
\hi\ masses less than $\mhis=6\times 10^9\,\msol$. The median $z=0$
DLA arises in a $L_*/7$ galaxy with an \hi\ mass of $2\times 10^9
\msol$.

3) The distribution of impact parameters and column densities agrees
very well between local galaxies and low-$z$ DLA galaxies. The median
impact parameter between the line of sight to a QSO and the centre of
the galaxy giving rise to a DLA is 7.8 kpc. For systems at $z=0.5$
($z=1$) we expect that $37$ per cent ($48$ per cent) have impact parameters less
than $1''$. These findings support indications that optical surveys for
DLA host galaxies miss identifications at very small impact
parameters, because of the brightness of the QSO or because of
blending due to too low spatial resolution. If obscuration of 
background QSOs by dust in DLA galaxies is important, this might
also have the strongest effect at small impact parameters.

4) We combine our data set with the well-established
luminosity-metallicity relation of galaxies and observed metallicity
gradients in galaxy disks to estimate the expected metallicity
distribution of low-$z$ DLAs. We find that the expected median
metallicity of $z=0$ DLAs is approximately $1/7$ solar, in good
agreement with observations of metal lines in DLAs.
The mean mass-weighted metallicity of the interstellar matter
in local galaxies above the DLA limit  is approximately half solar.
This is consistent with extrapolations from higher redshift measurements,
although the $z=0$ extrapolated value has large uncertainties given the
poor statistics from DLAs with redshifts approximately $z\approx1.5$.

5) The column density distribution function \fnhi\ in the local
Universe can be fitted satisfactorily with a gamma distribution. A
single power law is not a good fit. There is remarkably little
evolution in the shape of \fnhi\ from high $z$ to the present.

6) Most ($\approx 81$ per cent) of the cosmological mass density in \hi\ at
$z=0$ is locked up in column densities above the classical DLA limit
of $\nhi > 2 \times 10^{20}~\icmsq$, the rest is mostly in column
densities just below this limit.  The fraction is consistent over the
redshift range $z\sim 5$ to $z=0$.

\section*{Acknowledgments}
This work benefited from discussions with Nicolas Bouch{\'e},
C{\'e}line P{\'e}roux, Mike Fall, Palle M{\o}ller, and Jochen Liske on 
the nature of DLAs. We also thank them and the anonymous referee for useful 
comments on the manuscript. 
We are grateful to Sandhya Rao 
and Jason Prochaska 
for communicating results before publication.  
JMvdH acknowledges support from ATNF for
a distinguished visitor grant.  The Westerbork Synthesis Radio
Telescope is operated by The Netherlands Foundation for Research in
Astronomy (ASTRON) with financial support from The Netherlands
Organisation for Scientific Research (NWO).  We have made use of the
LEDA database (http://leda.univ-lyon1.fr). This research has made use
of the NASA/IPAC Extragalactic Database (NED) which is operated by the
Jet Propulsion Laboratory, California Institute of Technology, under
contract with the National Aeronautics and Space Administration.

\small
\bibliographystyle{mn2e}
\bibliography{mn-jour,all}

\label{lastpage}

\end{document}